\documentclass[journal]{IEEEtran}
\usepackage{amssymb,amsmath}
\usepackage{amsmath}
\usepackage{amsfonts}

\ifCLASSINFOpdf

\else

\fi

\usepackage{amssymb}
\usepackage{amsfonts}
\usepackage{latexsym,float}
\usepackage{graphicx}

\usepackage{color}

\usepackage{setspace}

\newcommand{\e}{\mbox{$\mathbb E$}}

\newcommand{\df}{\mbox{${\cal D}$}}

\newcommand{\n}{\mbox{$\cal N$}}

\newcommand{\oo}{\mbox{$\mathbb O$}}

\newcommand{\uu}{\mbox{$\mathbf u$}}
\newcommand{\x}{\mbox{$\mathbf x$}}
\newcommand{\y}{\mbox{$\mathbf y$}}
\newcommand{\vv}{\mbox{$\mathbf v$}}
\newcommand{\w}{\mbox{$\mathbf w$}}
\newcommand{\zz}{\mbox{$\mathbf z$}}

\newcommand{\aaa}{\mbox{$\cal A$}}
\newcommand{\ttt}{\mbox{$\cal T$}}
\newcommand{\f}{\mbox{$\cal F$}}
\newcommand{\bb}{\mbox{$\cal B$}}
\newcommand{\p}{\mbox{$\cal P$}}

\newcommand{\sss}{\mbox{$\cal S$}}
\newcommand{\rrr}{\mbox{$\cal R$}}

\newcommand{\up}{\mbox{$\it\Upsilon$}}

\newcommand{\po}{\mbox{$\it\Phi$}}

\newcommand{\rank}{\mathrm{rank\;}}

\newcommand{\qa}{\mbox{\quad\mbox{and}\quad}}

\newcommand{\bfX}{\mbox{\boldmath $X$}}

\newcommand{\bfP}{\mbox{\boldmath $P$}}

\newcommand{\rt}{\mbox{$\mathbb R$}}

 \def\diag{\mathop{{\rm diag}}\nolimits}

\newcommand{\bxi}{\mbox{\boldmath $\xi$}}

\newcommand{\eet}{\mbox{\boldmath $\eta$}}

\newtheorem{theorem}{Theorem}

\newtheorem{lemma}{Lemma}

\newtheorem{example}{Example}

\newtheorem{definition}{Definition}
\newtheorem{remark}{Remark}

\begin{document}
%
\title{Second Degree Model for Multi-Compression  and Recovery of Distributed  Signals}
%
%
%

\author{Pablo~Soto-Quiros, Anatoli~Torokhti and Stanley J. Miklavcic%
\thanks{Pablo Soto-Quiros is with the Centre for Industrial and Applied Mathematics, University of South Australia, SA 5095, Australia and Instituto Tecnologico de Costa Rica, Apdo. 159-7050, Cartago, Costa Rica (e-mail: juan.soto-quiros@mymail.unisa.edu.au).}
\thanks{Anatoli Torokhti is with the Centre for Industrial and Applied Mathematics, University of South Australia, SA 5095, Australia (e-mail: anatoli.torokhti@unisa.edu.au).}
\thanks{Stanley J. Miklavcic is with the Phenomics and Bioinformatics Research Centre, University of South Australia, SA 5095, Australia (e-mail: stan.miklavcic@unisa.edu.au).}
\thanks{Manuscript received XXXX XX, 2015; revised XXXX XX, 2015.}}


\markboth{}%
{Torokhti \MakeLowercase{\textit{et al.}}:  Multi-Compression  and Recovery of Distributed  Signals}

\maketitle

\begin{abstract}
We study the problem of multi-compression and reconstructing a stochastic signal  observed by  several independent  sensors (or compressors) that transmit compressed information to a fusion center. { The key aspect  of this  problem is to find models of the sensors and fusion center that  are optimized in the sense of an error minimization under a certain criterion, such as the mean square error (MSE).} { A novel technique to solve this problem is developed. The novelty is as follows. First,   the  multi-compressors are non-linear and   modeled using second  degree polynomials. This may increase the accuracy of the signal estimation   through the optimization in a higher dimensional parameter space compared to the linear case. Second, the required models are determined by a method based on a combination of the second degree transform (SDT) \cite{2001309} with the  maximum block improvement  (MBI) method \cite{chen87,Zhening2015} and the generalized rank-constrained matrix approximation \cite{Torokhti2007,tor5277}. It allows us to use the advantages of the methods in \cite{chen87,Zhening2015,Torokhti2007,tor5277} to further increase the estimation accuracy of the source signal.
Third, the proposed method is justified in terms of pseudo-inverse matrices. As a result, the models of compressors and fusion center always exist and  are numerically stable.} In other words, the proposed models may provide compression, de-noising and reconstruction of distributed signals in cases  when known  methods either are not applicable   or may produce  larger associated errors.

\end{abstract}



\IEEEpeerreviewmaketitle

\section{Introduction}

This paper deals with a new method for the multi-compression and recovery of a source  stochastic signal  for a non-linear  wireless sensor network (WSN). Although WSNs exhibit a significant potential in many application fields (see, for example \cite{Akyildiz20021}), up till now the results obtained are based on linear models and fall short of a consideration of more effective non-linear models for WSNs. The objective of this paper is to provide a new effective technique for the modeling of a non-linear  WSN. The technique is  based on an extension of the approach in \cite{2001309} in combination with ideas of the  maximum block improvement  (MBI) method \cite{chen87,Zhening2015} and  the generalized rank-constrained matrix approximation \cite{Torokhti2007,tor5277}.


\subsection{Motivation}

 \IEEEPARstart{W}{e} are motivated and inspired by the work in \cite{4276987,Song20052131, 1420805, 4016296,  4475387,Schizas2007, dragotti2009, 5447742, 5504834, Saghri2010, 6334203} where the effective methods for modeling of WSNs were developed.
  The WSNs are widely used in many areas of signal processing such as, for example, battlefield surveillance,
target localization and tracking, and acoustic beamforming using an array of microphones \cite{Li20021729,Lin2002102,Gastpar2004142} .
  An associated scenario involves a set of spatially distributed sensors, $\mathcal{B}_1,\ldots,\mathcal{B}_p$, and a fusion center $\ttt$. The sensors are autonomous, have a limited energy supply and  furthermore,  cannot communicate with each other.
The sensors make local noisy observations, ${\bf y}_1,\ldots,{\bf y}_p$, correlated with a source stochastic signal ${\bf x}$. Each sensor  $\mathcal{B}_j$ transmits compressed information about its
measurements, ${\bf u}_j$, to the fusion center which should recover the original signal
within a prescribed accuracy. The compression level is predefined and is not data-dependent.

The problem is  to find an effective way to {compress} and denoise each observation
${\bf y}_j$ , where $j = 1,\ldots, p$, and then {reconstruct} all the compressed observations in the
fusion center so that the reconstruction will be optimal in the sense of a minimization
of the associated error under a certain criterion, such as the mean square error (MSE).

{ The known methods for the  distributed signal compression and recovery \cite{4276987}-\cite{6334203} are based on a natural idea of extensions of the optimal {\em linear} transforms \cite{Brillinger2001,kailath2000,Poor2001}. In many cases, this approach leads to  a required associated accuracy. At the same time, the error associated with the optimal {\em linear} transform, for a given compression ratio, cannot be diminished by any other linear transform.
Thus, if the performance of the methods based on the optimal {\em linear} transforms  is not as good as required then  a method based on a different idea should be applied. In this paper, such a method is provided.

 The key questions motivating this work are as follows. How can one improve the accuracy of the distributed signal processing approach compared to that of other methods? Second, is it possible to achieve the improvement under the same assumptions that those in the known methods? The last question is motivated by an observation that the  models with an ``extended'' structure often require additional  initial information needed for their  implementation.
{ Third, the block coordinate descent (BCD) method \cite{Tseng2001} is known to work on the modeling of WSNs when applied only under certain conditions, which may restrict its applicability. Is it possible to avoid those restrictions?

}

\subsection{ Known techniques}\label{known}

Here, the term ``compression'' is treated in the same sense as in the previous works on data compression
(developed, for instance, in \cite{4276987}-\cite{6334203}, \cite{905856,Torokhti20102822}), i.e. we say that observed signal ${\bf y}_j$ with $n_j$ components is compressed if it is represented by signal ${\bf u}_j$ with $r_j$ components where $r_j < n_j$,
for $j = 1,\ldots, p$. That is ``compression'' refers to dimensionality reduction and not quantization
which outputs bits for digital transmission. This is similar to what is considered, in particular, in \cite{4276987,Song20052131, 1420805, 4016296, 4475387}.

 It is known that in the nondisrtibuted setting (in the other words,
in the case of a single sensor only) the MSE optimal solution is provided by the generic Karhunen-Lo\`{e}ve transform  (GKLT) \cite{2001309,torbook2007} { which provides the smallest associated error in the class of all {\em linear} transforms}. Nevertheless, the  GKLT cannot be applied to the above
WSN since the entire data vector ${\bf y} = [{\bf y}_1^T ,\ldots, {\bf y}_p^T]^T$ is not observed by each sensor (in this regard, see also \cite{4016296, 4475387}).
Therefore, several approaches to a determination of mathematical models for $\mathcal{B}_1 ,\ldots, \mathcal{B}_p$
and $\ttt$ have been pursued. In particular, in the information-theoretic context, distributed
compression has been considered in the landmark works of Slepian and Wolf \cite{1055037}, and
Wyner and Ziv \cite{1055508}.  { The natural setting of the approach in  \cite{1055037,1055508} in terms of  the transform-based methodology}  has been considered in \cite{4276987,Song20052131, 1420805, 4016296, 4475387}. Intimate relations
between these two perspectives have been shown in \cite{952802}. The methodology developed in
\cite{4276987,Song20052131, 1420805, 4016296,  4475387} is based on the dimensionality reduction by linear projections. Such an approach has received considerable attention (see, for example, \cite{Schizas2007, dragotti2009, 5447742, 5504834, Saghri2010, 6334203,Bert2010_1,Bert2010_2}).

In particular, in \cite{4276987}, two approaches are considered. By the first approach, the
fusion center model, $\ttt$, is given in the form $\ttt = [\ttt_1,\ldots,\ttt_p]$ where $\ttt_j$ is a `block' of
$\ttt$, for $j = 1,\ldots, p$, and then the original MSE cost function is represented as a sum
of $p$ decoupled MSE cost functions. Then approximations to $\mathcal{B}_j$ and $\ttt_j$ are found as
solution to each of the $p$ associated MSE minimization (MSEM) problems.
Some more related results can be found in \cite{Torokhti2012}.
 The second approach in \cite{4276987} generalizes the results in \cite{Song20052131, 1420805}
in the following way. The original MSE cost function is represented, by re-grouping
terms, in the form similar to that presented by the summand in the decoupled MSE cost
function. Then the minimum is seeking for each $\ttt_j\mathcal{B}_j$, for $j = 1,\ldots,p$, while other
terms $\ttt_k\mathcal{B}_k$, for $k = 1,\ldots,j - 1, j + 1,\ldots, p$, are assumed to be fixed. The minimizer
follows from the result given in \cite{Brillinger2001} (Theorem 10.2.4).
To combine solutions of those $p$ local MSEM problems, approximations to each sensor model $\mathcal{B}_j$ are determined from an iterative procedure. Values of $\mathcal{B}_1,\ldots,\mathcal{B}_p$ for
the initial iteration are chosen randomly.

The method in \cite{4475387} is based on ideas similar to those in the earlier references \cite{4276987,Song20052131, 1420805,4016296}\footnote{In particular,
it generalizes the work of \cite{4016296} to the case when the vectors of interest are not directly observed by the sensors.}, i.e., on the replacement of the original MSEM problem  with the $p + 1$ unconstrained MSEM
problems for separate determination of approximations to $\mathcal{B}_j$ , for each $j=1,\ldots, p$, and
then an approximation to $\ttt$. First, an approximation to each $\mathcal{B}_j$ , for $j = 1,\ldots, p$, is
determined under assumption that other $p-1$ sensors are fixed. Then, on the basis of
known approximations to $\mathcal{B}_1,\ldots,\mathcal{B}_p$, an approximation of $\ttt$ is determined as the optimal
Wiener filter.
In \cite{4475387}, the involved signals are assumed to be zero-mean jointly Gaussian
random vectors. Here, this restriction is not used.

The work in \cite{Song20052131, 1420805,4016296} can be considered as a particular case of \cite{4475387}.

 The method in \cite{ma3098} is applied to the problem which is an approximation of the original problem. It implies an increase in the associated error  compared to the method applied to the original problem. Further, the technique suggested in \cite{ma3098}  is applicable under certain restrictions imposed on  observations and associated covariance matrices. In particular, in \cite{ma3098}, the observations should be  presented in the special form ${\bf y}_j = H_j \x + \vv_j$, for $j=1,\ldots,p$ (where $H_j$ is a measurement matrix and $\vv_j$ is noise),  and the covariance matrix formed by the noise vector should be {\em block-diagonal} and invertible. It is not the case here.

The approaches taken in the above references are based, in fact, on the BCD method \cite{Tseng2001}. The BCD method  converges to a stationary point if the space of optimization is convex, and the objective function is continuous and regular  \cite{Tseng2001} (pp. 480--481). It converges to  a coordinatewise minimum if the space of optimization is convex and the  objective function is continuous   \cite{Tseng2001}  (pp. 480--481). The BCD method  converges to a global minimum if the objective function can be separated into sums of functions of single block variables \cite{Zhening2015} (p. 211). The above conditions are not satisfied in the present case.

Further, the WSN models in \cite{4276987,Song20052131, 1420805, 4016296,  4475387,ma3098} are  justified  in terms of inverse matrices.
It is known that in  cases when the matrices are close to singular this may lead to instability and a
significant increase in the associated error. Moreover, when the matrices are singular,
the algorithms \cite{4276987,Song20052131, 1420805, 4016296, 4475387} may not be applicable.
This observation is illustrated by Examples \ref{ex2}, \ref{ex4} in Section \ref{sim} and Example \ref{ex41} in Section \ref{sim2}  below where  the associated matrices are singular and the method  \cite{4276987} is not applicable.
 In \cite{4475387}, for the case when a matrix is singular, the inverse is replaced with the  pseudo-inverse,  but such a simple  replacement
does not follow from the justification of the  model provided in \cite{4475387}. As a result, a simple substitution by pseudo-inverse matrices  may result in  numerical instability as  shown, in particular, in \cite{grant2015} and   Example \ref{ex2}  in Section \ref{sim}.
In this regard, we also refer to references \cite{Torokhti2007,torbook2007,Torokhti2009661} where the case of rank-constrained data compression in terms of the pseudo-inverse matrices  is studied.


\subsection{ Differences from known methods. Novelty and Contribution}\label{differences}

 The  methods of \cite{4276987}-\cite{6334203},  \cite{ma3098}  are based on exploiting the {\em linear} minimizers of the mean square error   \cite{Brillinger2001,kailath2000,Poor2001} in  the BCD method \cite{Tseng2001}.  The key advantages and novelty  of the proposed methodology, and differences from the techniques in  \cite{4276987}-\cite{6334203},  \cite{ma3098} are as follows.

 Firstly, the minimizers developed in this paper are {\em non-linear}. They are based on the second degree transform (SDT) considered in \cite{2001309}.  It has been shown in \cite{2001309} that under a certain condition, the SDT leads to greater accuracy in the  signal estimation compared to  methods based on the optimal linear signal transform, the GKLT. In Section \ref{comparison lin}, a similar condition is determined for the case under consideration, i.e., for distributed signal compression and reconstruction.
Secondly,  unlike the previous works we apply a version of the maximum block improvement  (MBI) method  \cite{chen87,Zhening2015}, not the BCD method \cite{Tseng2001}. This is because  a solution of the minimization problem we consider is not unique,  the optimization space is not convex and the objective function is not regular in the sense  of \cite{Tseng2001}. The objective function also does not separate  into sums of functions of single block variables  \cite{Zhening2015}.
As a result, convergence of the BCD method  cannot be guaranteed \cite{Zhening2015} for the problem we consider. The MBI method \cite{chen87,Zhening2015} avoids the requirements of the BCD method. 
  The main idea of the MBI method is to implement an update of the block of variables which is the best possible among all the updates.
 Further, unlike the previous methods, at each stage of the proposed greedy method   the generalized rank-constrained matrix approximation \cite{Torokhti2007,tor5277} has to be applied. This is required because of  the special structure of the objective function under consideration. More details are given below in Section \ref{detp1pp}.
{Thirdly}, the minimizers in  \cite{Brillinger2001,kailath2000,Poor2001} used  in \cite{4276987}-\cite{6334203},  \cite{ma3098} have been obtained  in terms of inverse matrices, i.e., they have only been {\em justified} for full rank covariance matrices. In our method, the minimizers are obtained and rigorously {\em justified}  in terms of pseudo-inverse matrices and, therefore, the  models of the sensors and the fusion center are
numerically stable and always exist. In other words, the proposed WSN  model may provide compression, de-noising and reconstruction of distributed signals for the cases  when known  methods either are not applicable (because of singularity of associated matrices)  or produce  larger associated errors. This observation is supported, in particular, by the error analysis provided in Section \ref{convergence}  and by the results of simulations shown in Section \ref{sim}.
  Fourthly,   despite the special structure of our method compared to those of known techniques, the same initial information   is required for the method  implementation. This observation  is illustrated in Example \ref{ex1} given in Section \ref{sim}.

\begin{figure}[t]
\centering
\begin{tabular}{c}
\includegraphics[scale=0.27]{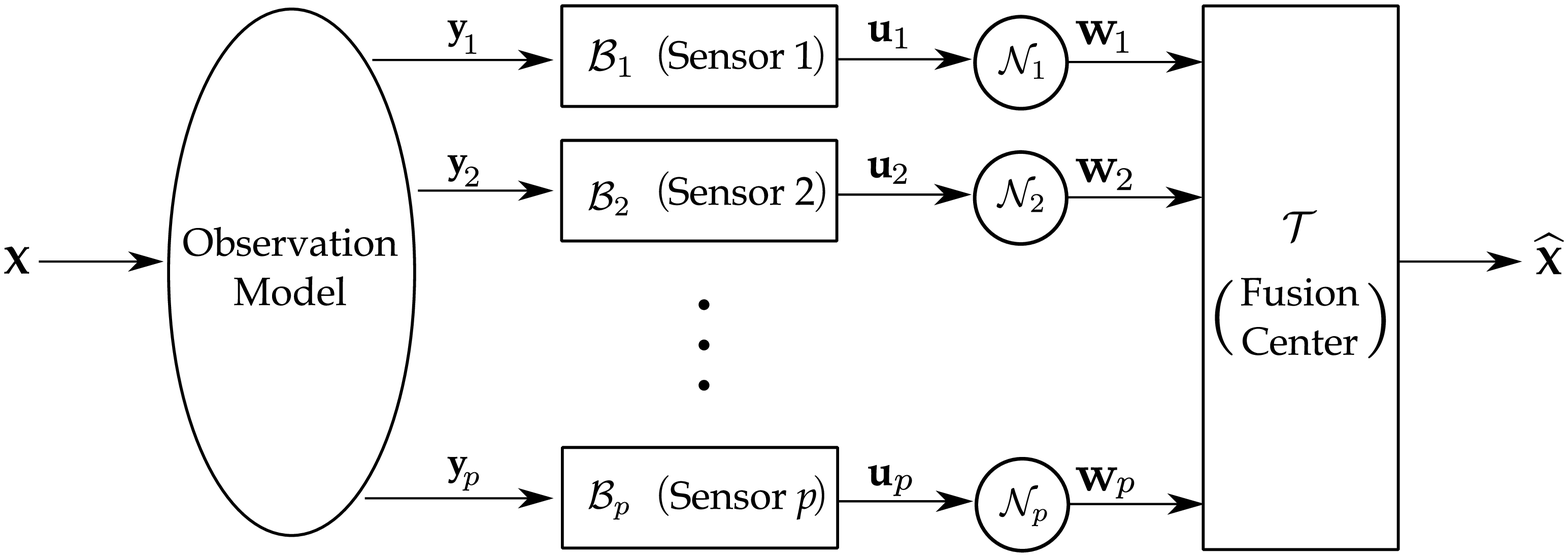}\\
(a)\\
\\
\includegraphics[scale=0.35]{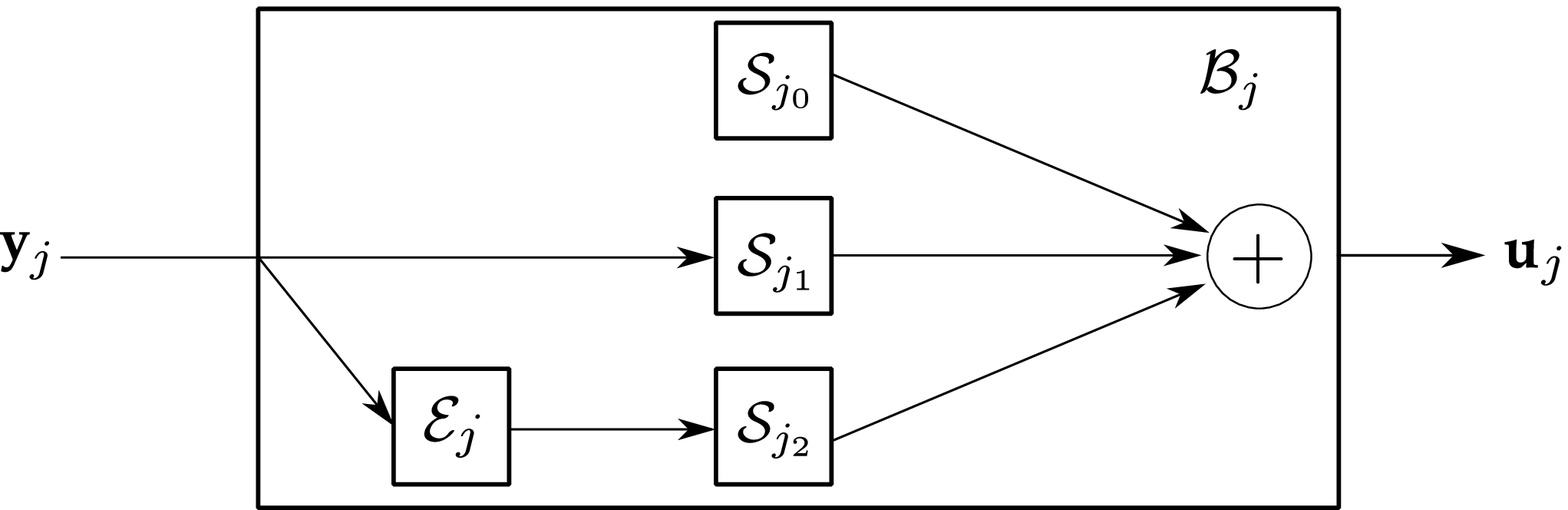}\\
(b)
\end{tabular}
\caption{(a): WSN model. (b): Sensor model. }
\label{fig1}
\end{figure}

\subsection{Notation}\label{notation}

 Here, we provide some notation which is required to formalize the problem
in the form presented in Section \label{statement_problem} below. Let us write $(\Omega,\Sigma,\mu)$ for a probability space\footnote{Here $\Omega=\{\omega\}$ is the set of outcomes, $\Sigma$ a $\sigma-$field of measurable subsets of $\Omega$ and $\mu:\Sigma\rightarrow[0,1]$ an associated probability measure on $\Sigma$ with $\mu(\Omega)=1$.}. We denote by ${\bf x} \in  L^2(\Omega,\mathbb{R}^m)$ the signal of interest\footnote{Space $L^2(\Omega,\mathbb{R}^m)$ has to be used because of the norm introduced in (\ref{eq6}) below.} (a source signal to be estimated) represented as ${\bf x}=[{\bf x}^{(1)},\ldots,{\bf x}^{(m)}]^T$ where ${\bf x}^{(j)}\in L^2(\Omega,\mathbb{R})$, for $j=1,\ldots,m$. Further, ${\bf y}_1 \in  L^2(\Omega,\mathbb{R}^{n_1}),\ldots,{\bf y}_p \in  L^2(\Omega,\mathbb{R}^{n_p})$ are observations made by the sensors. In this regard, we write
\begin{equation}\label{eq1}
{\bf y}=[{\bf y}^{T}_1,\ldots,{\bf y}^{T}_p]^T\;\text{ and }\;{\bf y}=[{\bf y}^{(1)},\ldots,{\bf y}^{(n)}]^T
\end{equation}
where ${\bf y}^{(k)} \in L^2(\Omega,\mathbb{R})$, for $k = 1,\dots, n$. We would like to emphasize a difference between
${\bf y}_j$ and $\y^{(k)}$: in (\ref{eq1}), the observation ${\bf y}_j$, for $j = 1,\dots,p$, is a random vector itself
(i.e. ${\bf y}_j$ is a `piece' of  ${\bf y}$), and ${\bf y}^{(k)}$, for $k = 1,\ldots,n$, is
a random variable (i.e. ${\bf y}^{(k)}$ is an entry of $\y$). Therefore,   $\y_j = (\y^{(q_{j-1} +1)},\ldots, \y^{(q_{j})})^T$ where $q_j = q_{j-1} + n_j$, $j=1,\ldots,p$ and $q_0=0$.

For $j = 1,\ldots, p$, we represent a sensor model by linear operator $\mathcal{B}_j : L^2(\Omega,\mathbb{R}^{n_j}) \rightarrow L^2(\Omega,\mathbb{R}^{r_j})$ defined by matrix  $B_j\in\mathbb{R}^{r_j\times n_j}$  such that,\footnote{An explanation for the relation in  (\ref{eq3}) is given in Section \ref{Explanation}. } for $j=1,\ldots,p$,
\begin{equation}\label{eq3}
[\mathcal{B}_j({\bf y}_j)](\omega)=B_j[{\bf y}_j(\omega)].
\end{equation}
Here,
$ 
r=r_1+\ldots+r_p,\;\text{,}\;r_j\leq \min \{m, n_j\},
$ 
and $r_j$  is given and fixed for each $j$th sensor, for $j=1,\ldots,p$.
 Let us denote ${\bf u}_j=\mathcal{B}_j({\bf y}_j)$ and ${\bf u}=[{\bf u}^{T}_1,\ldots,{\bf u}^{T}_p]^T$, where for $j=1,\ldots,p$, vector ${\bf u}_j\in L^2(\Omega,\mathbb{R}^{r_j})$ represents the compressed observation transmitted by a $j$th sensor $\mathcal{B}_j$ to the fusion center $\ttt$.

A fusion center model is represented  by linear operator $\ttt : L^2(\Omega,\mathbb{R}^{r}) \rightarrow L^2(\Omega,\mathbb{R}^{m})$ defined by matrix $T\in\mathbb{R}^{m\times r}$ so that
$ 
[\ttt({\bf u})](\omega)=T[{\bf u}(\omega)],
$ 
where  ${\bf u}\in L^2(\Omega,\mathbb{R}^r)$. To state the problem in the next section, we also
denote
\begin{equation}\label{eq6}
\e(\|{\bf x}\|^2): = \|{\bf x}\|^2_\Omega: =\int_\Omega \|{\bf x}(\omega)\|_2^2 d\mu(\omega) < \infty,
\end{equation}
where $\|{\bf x}(\omega)\|_2$ is the Euclidean norm of ${\bf x}(\omega)\in\mathbb{R}^m$.

\section{  Statements of the  Problems } \label{preliminaries}

\subsection{Formalization of the Generic Problem} \label{discussion_problem}

For ${\bf x}$ represented by ${\bf x} = [{\bf x}^{(1)},\ldots, {\bf x}^{(m)}]^T$
 where $\x^{(j)}\in L^2(\Omega,\mathbb{R})$  we write
$ E[{\bf xy}^T]$ $=E_{xy}$ $=\left\{ E_{x_jy_k}\right\}_{j,k=1}^{m,n}\in\mathbb{R}^{m\times n},$
where $\displaystyle E_{x_jy_k}=\int_\Omega {\bf x}^{(j)}(\omega){\bf y}^{(k)}(\omega) d\mu(\omega) < \infty$ and ${\bf y} = [{\bf y}^{(1)},\ldots, {\bf y}^{(n)}]^T$.
An assumption used in previous methods \cite{ 4276987,Song20052131, 1420805, 4016296, 4475387} and associated works such as \cite{Brillinger2001,kailath2000,Poor2001,905856,tor199} is that the covariance matrices
$E_{xy}$ and $E_{yy}$  are known. Here, we adopt this assumption.   As mentioned, in particular, in \cite{4276987} (and similarly, for example, in \cite{tor199,149980,Ledoit2004365,ledoit2012,Adamczak2009,Vershynin2012,won2013,schmeiser1991,yang1994}), ``{\em a priori} knowledge of the covariances can come either from specific data models, or, after sample estimation during a training phase." Importantly, although  we wish to develop the method with an extended structure,  no  additional initial information is required.  Example \ref{ex1} in Section \ref{sim} illustrates this observation. 

First,  we consider the case when the channels from the sensors to the fusion center are ideal, i.e., each channel operator $\n_j$ in Fig. \ref{fig1} (a) is the identity. The case of not ideal channels will be considered in Section \ref{nonideal}.
Then the problem is as follows:
Find models of the sensors, $\mathcal{B}_1,\ldots,\mathcal{B}_p$, and a model of the fusion center, $\ttt$, that provide
\begin{equation}\label{eq7}
\min_{\mathcal{T},\mathcal{B}_1,\ldots,\mathcal{B}_p} \left\| \x - \ttt\left[ \begin{array}{c}
                                                                                  \mathcal{B}_1({\bf y}_1) \\
                                                                                   \vdots \\
                                                                                  \mathcal{B}_p({\bf y}_p)
                                                                                 \end{array}\right] \right\|_\Omega^2.
\end{equation}
The model of the fusion center, $\ttt$, can be represented as $\ttt = [\ttt_1,\ldots,\ttt_p]$ where for
$j = 1,\ldots,p$, $\ttt_j : L^2(\Omega,\mathbb{R}^{r_j}) \rightarrow L^2(\Omega,\mathbb{R}^{m})$.
Let us write
\begin{eqnarray}\label{xt1pb1bp1}
\hspace{-10cm}\min_{\substack{\mathcal{T}_1,\ldots, \mathcal{T}_p, \\ \mathcal{B}_1,\ldots,\mathcal{B}_p}}\left\| \x - [\ttt_1,\ldots,\ttt_p]\left[ \begin{array}{c}
                              \mathcal{B}_1({\bf y}_1) \\
                               \vdots \\
                               \mathcal{B}_p({\bf y}_p)
                             \end{array}\right] \right\|_\Omega^2\nonumber  \\
= \min_{\substack{\mathcal{T}_1,\ldots, \mathcal{T}_p, \\ \mathcal{B}_1,\ldots,\mathcal{B}_p}}\|{\bf x}-[\ttt_1\mathcal{B}_1({\bf y}_1)+\ldots+\ttt_p\mathcal{B}_p({\bf y}_p)]\|_\Omega^2 \label{eq9}
\end{eqnarray}
where $\y = [ {\bf y}_1^T,\ldots,  {\bf y}_p^T ]^T.$

In this paper, we study the case where $\bb_j$, for $j=1,\ldots,p$, is represented by a second degree polynomial, i.e., is non-linear.
Previous methods were developed for the  case of {\em linear} $\ttt_j$ and  $\bb_j$, for $j=1,\ldots,p$.



\subsection{Inducement to Introduce Second Degree WSN} \label{mot2}

\subsubsection{Linear WSN} \label{linear} Let us first consider the case when $\ttt_j$ and  $\bb_j$ are {\em linear} operators, and $\n_j=I$, for $j=1,\ldots,p$.
Denote by $\rrr(m, n_j, r_j)$ the variety of all  linear operators $L^2(\Omega,\mathbb{R}^{n_j}) \rightarrow L^2(\Omega,\mathbb{R}^{m})$ of rank  at most $r_j$ where $k\leq \min\{m, n_j\}.$ For convenience we will sometimes  write $\rrr_{r_j}$ instead of $\rrr(m, n_j, r_j)$.
Then, for $\f_j = \ttt_j \bb_j$,   the generic problem in (\ref{eq9}) can equivalently be reformulated as follows: Find $\f_1,\ldots \f_p$ that solve
\begin{equation}\label{f1p}
\min_{\mathcal{F}_1\in \mathcal{R}_{r_1},\ldots,\mathcal{F}_p\in\mathcal{R}_{r_p}} \left\|{\bf x}-\sum_{j=1}^p \f_j(\y_j)\right\|_\Omega^2.
\end{equation}
Recall that $r_j < \min\{m, n_j\}$, for $j=1,\ldots,p$.

\subsubsection{Second Degree WSN} Mathematically speaking, the problem in (\ref{f1p}) is the problem of the best constrained {\em linear} approximation.   At the same time, it is  known that a `proper' {\em non-linear} approximation would  provide   better associated  accuracy, i.e., a better estimation of $\x$.
This follows, in particular, from  the famous Weierstrass approximation theorem \cite{Kreyszig1978} and its generalizations in \cite{Prenter341,Istr118,Bruno13,Akhiezer1992,Sandberg40,Mathews2001,Howlett353}.\footnote{A constructive choice of the optimal non-linear approximation, which provides the minimal associated error, is a special and hard research problem \cite{Deutsch2001}. Its solution is subject to special conditions (convexity, for example) that are not satisfied for the problem under consideration. This is why in \cite{4276987,Song20052131, 1420805, 4016296, 4475387,Schizas2007, dragotti2009, 5447742}, the special approaches for solving problem (\ref{f1p}) have been developed.} In  \cite{2001309}, this observation has been realized as the second degree transform.  For $p=1$ and $\n_1=I$ in (\ref{xt1pb1bp1}), it has been shown in \cite{2001309}  that if
\begin{equation}\label{t1b1y1}
\ttt_1\mathcal{B}_1({\bf y}_1) = \aaa_1(\y) + \aaa_2 (\y^2),
\end{equation}
where   $\y^2$ is given by $\y^2(\omega) := ([\y_1(\omega)]^2,\ldots, [\y_n(\omega)]^2)^T$, and operators $\aaa_1$ and $ \aaa_2$ are determined from the minimization of the associated MSE,  then under a quite unrestrictive condition obtained in \cite{2001309}, the error associated with the SDT \cite{2001309} is less than the error associated with the GKLT, for the same compression ratio.
  The improvement in  \cite{2001309} is due to doubling of the number of parameters compared to the KLT. Indeed, the KLT  is a particular case of transform \cite{2001309} if  $\aaa_2=\oo$ in (\ref{t1b1y1}), where $\oo$ is zero operator.
In the general case,    optimization of  $\ttt_1\mathcal{B}_1$ in (\ref{t1b1y1})  is achieved by a variation of two operators, ${\aaa}_{1}$ and ${\aaa}_{2},$ while the KLT is obtained on the basis of optimization of ${\aaa}_1$ only. More details are provided in Section \ref{nonlinear}.

Based on these observations, we now extend the approach in \cite{2001309} to the case of arbitrary $p$ in (\ref{xt1pb1bp1}) and combine  it with the MBI method \cite{chen87,Zhening2015} and results in \cite{Torokhti2007,tor5277}. To this end, we write
 \begin{eqnarray}\label{qqqj}
\bb_j (\y_j) =  \sss_{j, 0} + \sss_{j, 1}(\y_j) + \sss_{j, 2}(\y^2_j),
\end{eqnarray}
where a random vector $\sss_{j, 0}\in L^2(\Omega,\mathbb{R}^{r_j})$, and  linear operators  $\sss_{j, 1}: L^2(\Omega,\mathbb{R}^{n_j}) \rightarrow L^2(\Omega,\mathbb{R}^{r_j})$ and $\sss_{j, 2}: L^2(\Omega,\mathbb{R}^{n_j}) \rightarrow L^2(\Omega,\mathbb{R}^{r_j})$ are to be determined, and $\y^2_j$ is defined by $\y^2_j(\omega) = ([\y^{(q_{j-1} +1)}(\omega)]^2,$ $\ldots, [\y^{(q_{j})}(\omega)]^2)^T$, for all $\omega\in\Omega$, where, as before, $q_j = q_{j-1} + n_j$, $j=1,\ldots,p$ and $q_0=0$. In (\ref{qqqj}), the term $\sss_{j, 2}(\y^2_j)$ can be interpreted as a {\em `second degree term'.} In this regard, $\bb_j (\y_j)$  is called the  second degree polynomial.
Furthermore,  $\bb_j (\y_j)$ can also be written as
 \begin{eqnarray}\label{qqqj2}
\bb_j (\y_j) = \sss_j (\zz_j)
\end{eqnarray}
where
\begin{eqnarray}\label{sssj02}
\sss_j = [\sss_{j, 0}, \sss_{j, 1}, \sss_{j, 2}] \qa \zz_j = [1, \y_j^T, (\y^2_j)^T]^T.
\end{eqnarray}
 The above implies the following definition.

 \begin{definition}\label{def1}
 Let
 $\sss = \mbox{diag} [\sss_1,\ldots \sss_p],$   $\zz = [\zz_1^T,\ldots,\zz_p^T]^T$ and $\p_j = \ttt_j \sss_j,$ for $j=1,\ldots,p$.
 The  WSN represented by the operator $\p = \ttt\sss$  such that
\begin{eqnarray}\label{ptsjyj}
 \p(\zz)  & = & \displaystyle \sum_{j=1}^p \p_j(\zz_j) \nonumber\\
          & = & \sum_{j=1}^p \ttt_j \left[\sss_{j, 0} + \sss_{j, 1}(\y_j) + \sss_{j, 2}(\y^2_j)\right],
\end{eqnarray}
where
\begin{eqnarray}\label{pjzzj1}
\p_j(\zz_j)  & = & \ttt_j \sss_j (\zz_j)\nonumber\\
             & = & \displaystyle \ttt_j \left[\sss_{j, 0} + \sss_{j, 1}(\y_j) + \sss_{j, 2}(\y^2_j)\right]
\end{eqnarray}
will be called the {\em second degree} WSN.  If $\sss_{j, 0}$ and  $\sss_{j, 2}$ are the zero random vector and zero operator, respectively, for all $j=1,\ldots,p$, then the WSN will be called the {\em linear} WSN.
 \end{definition}

\subsection{Statement of the Problem for Second Degree WSN} \label{stat2}

On the basis of (\ref{qqqj})-(\ref{ptsjyj}), for the second degree WSN, the generic problem in (\ref{eq9}) is reformulated as follows. For $j=1,\ldots,p$, let $\bb_j$ and $\zz_j$ be represented by (\ref{qqqj})-(\ref{sssj02}), and $\p_j$ be as in (\ref{pjzzj1}). Find  $\p_1,\ldots \p_p$ that solve
\begin{equation}\label{p1p}
\min_{\mathcal{P}_1\in\mathcal{R}_{r_1},\ldots,\mathcal{P}_p\in\mathcal{R}_{r_p}} \left\|{\bf x}-\sum_{j=1}^p \p_j(\zz_j)\right\|_\Omega^2.
\end{equation}
Note that $\p_j$ is linear with respect to $\zz_j$ and non-linear with respect to $\y_j$. At the same time, we are interested in the optimal models for $\bb_j$, for $j=1,\ldots,p$, and $\ttt$. It will be show in Section \ref{second degree}, that the models follow from the solution of problem (\ref{p1p}).



\section{ Solution of Problem (\ref{p1p})}\label{second degree}

\subsection{Greedy Approach to Determining $\mathcal{P}_1,\ldots,\mathcal{P}_p$ that Solve (\ref{p1p})}\label{detp1pp}

We wish to find $\mathcal{P}_1,\ldots,\mathcal{P}_p$ that provide a solution to the problem in  (\ref{p1p}) for an arbitrary finite number of sensors
in the  WSN model, i.e. for $p=1,2,\ldots$ in (\ref{p1p}). To this end, we need  some more preliminaries which are given in Sections \ref{reduction} and \ref{svdsvd} that follow. The method itself and an associated algorithm are then represented in  Section \ref{structure}.

\subsubsection{Reduction of Problem (\ref{p1p}) to Equivalent Form}\label{reduction}

Let us set
$M^{1/2\dagger}=(M^{1/2})^\dagger$, where $M^{1/2}$ is a square root of a matrix $M$, i.e. $M = M^{1/2}M^{1/2}$. The pseudo-inverse for a matrix $M$ is denoted by $M^\dagger$.

We denote $P=[P_1,\ldots,P_p]$, where $P\in\mathbb{R}^{m\times (2n+p)}$ and $P_j\in\mathbb{R}^{m\times (2n_j+1)}$, for all $j=1,\ldots,p$, $\p_j(\zz_j)=P_j\zz_j$
and write $\|\cdot\|$ for the Frobenius norm. Then
\begin{eqnarray}\label{xfy1p}
\left\|{\bf x}-\sum_{j=1}^p \mathcal{P}_j(\zz_j)\right\|_\Omega^2  & = \hspace{-0.1cm} & \left\|{\bf x}-\sum_{j=1}^p P_j\zz_j\right\|_\Omega^2\nonumber\\
                                                                   & = \hspace{-0.1cm} & \|{\bf x}-P({\bf z})\|_{\Omega}^2\nonumber\\
                                                                   & = \hspace{-0.1cm} & \|E_{xx}^{1/2}\|^2-\|E_{xz}(E_{zz}^{1/2})^\dagger\|^2\nonumber\\
                                                                   &   \hspace{-0.1cm} & +\| E_{xz}(E_{zz}^{1/2})^\dagger \hspace{-0.05cm} - \hspace{-0.05cm} PE_{zz}^{1/2} \|^2.
\end{eqnarray}
Let us write $H=E_{xz}(E_{zz}^{1/2})^\dagger$ and represent matrix $E_{zz}^{1/2}$ in blocks
\begin{equation}\label{ezzg1p}
E_{zz}^{1/2}= [G_{1}^T,\ldots,G_{p}^T ]^T
\end{equation}
where $E_{zz}^{1/2}\in \rt^{(2n+p)\times (2n+p)}$  and  $G_j\in \rt^{(2n_j+1)\times (2n+p)}$, for $j=1,\ldots,p$. Then
\begin{eqnarray}\label{pezzghj}
E_{xz}(E_{zz}^{1/2})^\dagger - PE_{zz}^{1/2}  = H - \sum_{j=1}^p P_jG_j.
\end{eqnarray}
Therefore, (\ref{xfy1p}) and (\ref{pezzghj}) imply
\begin{eqnarray}\label{p1ppr1p}
\min_{P_1\in {\mathcal R}_{r_1},\ldots, P_p\in {\mathcal R}_{r_p}} \left\|{\bf x}-\sum_{j=1}^p P_j(\zz_j)\right\|_\Omega^2& &\nonumber\\
=  \min_{P_1\in {\mathbb R}_{r_1},\ldots, P_p\in {\mathbb R}_{r_p}}\left\|H - \sum_{j=1}^p P_jG_j\right\|^2 & &
\end{eqnarray}
where  $\rt_{r_j} $ denotes the variety of all $m \times (2n_j+1)$ matrices of rank  at most $r_j$, for $j=1,\ldots,p$, where $r_j\leq \min\{m, n_j\}$. On the basis of (\ref{p1ppr1p}), problem (\ref{p1p}) and the problem
\begin{eqnarray}\label{p1ph1p}
 \min_{P_1\in {\mathbb R}_{r_1},\ldots, P_p\in {\mathbb R}_{r_1}}\left\|H - \sum_{j=1}^p P_jG_j \right\|^2
\end{eqnarray}
are equivalent. Therefore, below we consider problem  (\ref{p1ph1p}). Note that
 \begin{eqnarray}\label{hpjgj1}
\left\|H - \sum_{j=1}^p P_jG_j \right\|^2 = \left\|Q_j - P_jG_j\right\|^2,
\end{eqnarray}
where  $\displaystyle Q_j = H-\sum_{\substack{i=1\\i\neq j}}^pP_iG_i$. This representation will be used below.


\subsubsection{SVD and Orthogonal Projections }\label{svdsvd}   Let the SVD of a matrix $C \in \mathbb{R}^{m\times s}$ be given by
\begin{equation}\label{eq19}
C=U_C\Sigma_C V^T_C,
\end{equation}
where $U_C\in\mathbb{R}^{m\times m}$ and $V_C\in\mathbb{R}^{s\times s}$ are unitary matrices, $\Sigma_C=\text{diag}(\sigma_1(C),\ldots,\sigma_{\min(m,s)}(C))\in\mathbb{R}^{m\times s}$ is a generalized diagonal matrix, with the singular values $\sigma_1(C)\geq\sigma_2(C)\geq\ldots0$ on the main diagonal.
Let $U_C=[u_1\;u_2\;\ldots\;u_m]$ and $V_C=[v_1\;v_2\;\ldots\;v_s]$ be the representations of $U$ and $V$ in terms of their $m$ and $s$ columns, respectively. Let $L_C\in\mathbb{R}^{m\times m}$ and $R_C\in\mathbb{R}^{s\times s}$, such that
\begin{equation}\label{lcrc}
L_C=\sum_{i=1}^{\text{rank }C}u_iu_i^T\;\;\text{ and }\;\;R_C=\sum_{i=1}^{\text{rank }C}v_iv_i^T,
\end{equation}
be the orthogonal projections on the range of $C$ and $C^T$, correspondingly.  Define $C_r\in\mathbb{R}^{m\times s}$ such that
\begin{equation}\label{eq21}
C_r=[C]_r=\sum_{i=1}^r\sigma_i(C)u_iv_i^T=U_{C_r}\Sigma_{C_r}V_{C_r}^T
\end{equation}
for $r=1,\ldots,\text{rank }C$, where
\begin{equation*}
U_{C_r}=[u_1\;u_2\;\ldots\;u_r],
\end{equation*}
\begin{equation}\label{eq22}
\Sigma_{C_r}=\text{diag}(\sigma_1(C),\ldots,\sigma_r(C)),
\end{equation}
\begin{equation*}
V_{C_r}=[v_1\;v_2\;\ldots\;v_r].
\end{equation*}

For $r>\text{rank }C$, we write $C^{(r)}=C(=C_{\text{rank }C})$. For $1\leq r<\text{rank }C$, the matrix $C^{(r)}$ is uniquely defined if and only if $\sigma_r(C)>\sigma_{r+1}(C)$.

\subsubsection{Greedy Method for  Solution of Problem (\ref{p1ph1p})}\label{structure}

An exact solution of problem (\ref{p1ph1p}) is unknown. That is why, similar to previous works \cite{4276987}-\cite{6334203},  \cite{ma3098}, we adopt a greedy approach for its solution.\footnote{Other reason to apply the  greedy approach is mentioned at the end of Section \ref{nonlinear}.} In particular, in \cite{4276987}-\cite{6334203},   \cite{ma3098}, the BCD method   \cite{Tseng2001} (mentioned in Section \ref{known}) was used.
At the same time, the conditions for convergence of the BCD method  are not satisfied for problem (\ref{p1ph1p}).
The recently developed MBI method \cite{chen87,Zhening2015} avoids the restrictions of the BCD method. The advantage and main idea of the MBI method is that  it accepts an ``update of the block of variables that achieves the maximum improvement" \cite{Zhening2015} at each iteration.
 Further, the problem in (\ref{p1ph1p}) is different from that considered in \cite{4276987}-\cite{6334203},   \cite{ma3098}. Therefore, unlike methods in \cite{4276987}-\cite{6334203},  \cite{ma3098}, a solution of problem (\ref{p1ph1p}) is represented by a version of the  MBI method \cite{chen87,Zhening2015} where at each iteration the result from \cite{Torokhti2007,tor5277}   is exploited;
 this is due to the specific form of the objective function in  (\ref{p1ph1p}).  More precisely,  each iteration of the proposed method is based on the following result.

\begin{theorem}\label{th01}
Let  $K_j= M_j\left(I-L_{G_{j}}\right)$ where  $M_j$ is an arbitrary matrix. For given $P_1,\ldots,P_{j-1},$ $P_{j+1},$ $\ldots,P_p$, the optimal matrix $\widehat{P}_j$ of rank at most $r_j$  minimizing $$\left\|{\bf x}-\sum_{j=1}^p \p_j(\zz_j)\right\|_\Omega^2$$ is given by
\begin{equation}\label{fjhjrgj}
\widehat{P}_j=\left[Q_jR_{G_j}\right]_{r_j}G_j^\dagger (I+K_j), \quad \mbox{for $j=1,\ldots,p$.}
\end{equation}
where $Q_j$ and $G_j$ are as in (\ref{ezzg1p}) and (\ref{hpjgj1}),  and   $R_{G_j}$ and  $[\cdot]_{r_j}$  are defined similarly to  (\ref{lcrc}) and  (\ref{eq21}), respectively. Any minimizing $\widehat{P}_j$ has the
above form if and only if either
$
r_j\geq \text{rank }(Q_jR_{G_j})
$
or
$
1\leq r_j<\text{rank }(Q_jR_{G_j})\;\;\text{ and }\sigma_{r_j}(Q_jR_{G_j})>\sigma_{{r_j}+1}(Q_jR_{G_j}),
$
 where $\sigma_{r_j}(Q_jR_{G_j})$ is a singular value in the SVD for matrix $Q_jR_{G_j}$.
\end{theorem}
\begin{IEEEproof}
The proof is considered in Section \ref{convergence}. We note that the arbitrary matrix $M_j$ implies non-uniqueness of  the optimal matrix  $\widehat{P}_j$.
\end{IEEEproof}

{We represent the error  associated with the optimal matrix $\widehat{P}_j$ in the way similar to that used in \cite{4276987,4475387}.} Let us write $\overline{\x}=\x- \w_j,$ where $\w_j=\sum_{i=1 \atop i\neq j}^p\p_i\zz_i$.Denote by
$\delta_1,...,\delta_{r_j}$  the first $r_j$ eigenvalues in the SVD for matrix $E_{\overline{x}z_j}E_{z_jz_j}^\dagger E_{z_j\overline{x}}$,
and by  $\mu_{j,1},...,\mu_{j,m_j}$  the eigenvalues of $C_jH_j^\dagger C_j^T$, where
$C_j=E_{\overline{x}y_j^2}-E_{\overline{x}y_j}E_{y_jy_j}^\dagger E_{y_jy_j^2}$ and $H_j=E_{y_j^2y_j^2}-E_{y_j^2y_j}E_{y_jy_j}^\dagger E_{y_jy_j^2}$.
We also write $\beta_j=\text{tr}\{2\;E_{w_jx}-E_{w_jw_j}\}$
and  $$\left.f(P_1,...,P_p) = \left\|{\bf x}-\sum_{j=1}^p \p_j(\zz_j)\right\|_\Omega^2\right..$$  

\begin{theorem}\label {th011}
 For given $P_1,\ldots,P_{j-1},$ $P_{j+1},$ $\ldots,P_p$, the error associated with the optimal matrix $\widehat{P}_j$ is given by
\begin{equation}\label{errorf1p}
f(P_1,...,P_{j-1},\widehat{P}_j,P_{j+1},...,P_p) =  \text{tr}\{E_{xx}\}-\sum_{i=1}^{r_j}\delta_i-\sum_{i=1}^{m}\mu_{j,i}-\beta_j.
\end{equation}
\end{theorem}
\begin{IEEEproof}
The proof is considered in Section \ref{convergence}. If the given matrices  $P_1,\ldots,P_{j-1},$ $P_{j+1},$ $\ldots,P_p$ are optimal in the sense of minimizing $\left\|{\bf x}-\sum_{j=1}^p \p_j(\zz_j)\right\|_\Omega^2$  then Theorem  \ref{th01} returns the globally optimal solution of (\ref{p1p}).

\end{IEEEproof}

Further, let us denote $\rt_{r_1,\ldots,r_p} = \rt_{r_1}\times...\times\rt_{r_p},$ $\quad{\bfP}=\left(P_1,...,P_p\right)\in\rt_{r_1,\ldots,r_p}$ and $\phi({\bfP}) = \left\|H - \sum_{j=1}^p P_jG_j \right\|^2.$ The computational procedure for the solution of problem (\ref{p1p}) is based on the solution of problem (\ref{p1ph1p}) and consists of the following steps.

{\em $1$st step.}  Given ${\bfP}^{(0)}=(P^{(0)}_1,...,P^{(0)}_p)\in\rt_{r_1,\ldots,r_p},$ compute $\widehat{P}^{(1)}_j$, for $j=1,\ldots,p$, such that
 \begin{eqnarray}\label{whf1j}
\widehat{P}^{(1)}_j = \left[Q^{(0)}_jR_{G_j}\right]_{r_j}G_j^\dagger (I+K_j),
\end{eqnarray}
where $$\displaystyle Q^{(0)}_j= H-\sum_{\substack{i=1\\i\neq j}}^pP^{(0)}_iG_i.$$
Here,  $G_j$ is given by (\ref{hpjgj1}), and $\widehat{P}^{(1)}_j$ is evaluated in the form (\ref{whf1j})  on the basis of Theorem \ref{th01} and using $P^{(0)}_1,...,P^{(0)}_{j-1}, P^{(0)}_{j+1},$ $\ldots,$ $P^{(0)}_p$.
Matrices $E_{xz}$ and $E_{zz}$ are formed from matrices $E_{xy}$ and $E_{yy}$. This is illustrated below by Example \ref{ex1} in Section \ref{sim}. A choice of initial iterations $P^{(0)}_1,\ldots, P^{(0)}_p,$ is considered in Section \ref{initial_iterations} below.

{\em $2$nd step.} Denote
  \begin{eqnarray}\label{bf1j}
  \overline{\bfP}^{(1)}_j = \left(P^{(0)}_1,\ldots,P^{(0)}_{j-1},\widehat{P}_j^{(1)},P^{(0)}_{j+1},\ldots,P^{(0)}_{p}\right)
\end{eqnarray}
and select $\overline{\bfP}^{(1)}_k$, for $k=1,\ldots,p$, such that  $\phi(\overline{\bfP}^{(1)}_k)$ is minimal among all $\phi(\overline{\bfP}^{(1)}_1), \ldots, \phi(\overline{\bfP}^{(1)}_k),\ldots, \phi(\overline{\bfP}^{(1)}_p)$, i.e.
\begin{eqnarray}\label{bf1k}
  \overline{\bfP}^{(1)}_k = \arg \min_{\mbox{\scriptsize\boldmath $\overline{P}$}^{(1)}_1,\ldots,\mbox{\scriptsize\boldmath $\overline{P}$}^{(1)}_p} \left \{\phi(\overline{\bfP}^{(1)}_1), \ldots, \phi(\overline{\bfP}^{(1)}_k),\ldots, \phi(\overline{\bfP}^{(1)}_p) \right\}
\end{eqnarray}
and write
\begin{eqnarray}\label{bff1}
{\bfP}^{(1)} = \overline{\bfP}^{(1)}_k,
\end{eqnarray}
where we denote ${\bfP}^{(1)} = (P^{(1)}_1,...,P^{(1)}_p)\in\rt_{r_1,\ldots,r_p}$.

Then we repeat procedure (\ref{whf1j})-(\ref{bff1}) with the replacement of ${\bfP}^{(0)}$ by ${\bfP}^{(1)}$  as follows: Given ${\bfP}^{(1)} = (P^{(1)}_1,...,P^{(1)}_p),$ compute, for $j=1,\ldots,p,$
$$\widehat{P}^{(2)}_j = \left[Q^{(1)}_jR_{G_j}\right]_{r_j}G_j^\dagger (I+K_j),$$
where $$\displaystyle Q^{(1)}_j= H-\sum_{\substack{i=1\\i\neq j}}^pP^{(1)}_iG_i.$$
Note that $\widehat{P}^{(2)}_j$ is computed similar to $\widehat{P}^{(1)}_j$, i.e., on the basis of Theorem \ref{th01} and using  $P^{(1)}_1,...,P^{(1)}_{j-1}, P^{(1)}_{j+1},$ $\ldots,$ $P^{(1)}_{p}$.
Then denote $ \overline{\bfP}^{(2)}_j = \left(P^{(1)}_1,\ldots,P^{(1)}_{j-1},\widehat{P}_j^{(2)},P^{(1)}_{j+1},\ldots,P^{(1)}_{p}\right)$, select $\overline{\bfP}^{(2)}_k$ that satisfies (\ref{bf1k}) where superscript $(1)$ is replaced with  superscript $(2)$, and set  ${\bfP}^{(2)} = \overline{\bfP}^{(2)}_k$ where ${\bfP}^{(2)} = (P^{(2)}_1,...,P^{(2)}_p)\in\rt_{r_1,\ldots,r_p}$.

This process is continued up to the $q$th step when a given tolerance $\epsilon \geq 0$  is achieved in the sense
\begin{eqnarray}\label{qeps1}
|\phi({\bfP}^{(q+1)}) - \phi({\bfP}^{(q)}) |\leq \epsilon, \quad \mbox{for $q =1,2,\ldots$}.
\end{eqnarray}
It is summarized as follows.

\vspace{5mm}

\hrule

\vspace{0.1cm}

\noindent{\bf Algorithm 1:} Greedy solution of problem  (\ref{p1ph1p})

\vspace{0.1cm}

\hrule

\vspace{0.1cm}

\noindent{\bf Initialization:} ${\bfP}^{(0)}$,
$H$,  $G_1,...,G_p$ and $\epsilon>0$.

\vspace{0.25cm}

\hrule

\begin{enumerate}

\item {\bf for }$q=0,1,2,...$

\item \hspace{0.1cm} {\bf for }$j=1,2,...,p$

\item \hspace{0.3cm} $Q^{(q)}_j= H-\sum_{\substack{i=1,\;i\neq j}}^pP^{(q)}_iG_i$

\item \hspace{0.3cm} $\widehat{P}^{(q+1)}_j = \left[Q^{(q)}_jR_{G_j}\right]_{r_j}G_j^\dagger (I+K_j)$

\item \hspace{0.3cm} $\overline{\bfP}^{(q+1)}_j = \left(P^{(q)}_1,...,P^{(q)}_{j-1},\widehat{P}_j^{(q+1)},P^{(q)}_{j+1},...,P^{(q)}_{p}\right)$

\item \hspace{0.1cm} {\bf end}

\item \hspace{0.1cm} Choose $\overline{\bfP}^{(q+1)}_k$  such that
$$\displaystyle\hspace{-0.3cm}\overline{\bfP}^{(q+1)}_k=\arg \min_{\mbox{\scriptsize\boldmath $\overline{P}$}^{(q+1)}_1,\ldots , \mbox{\scriptsize\boldmath $\overline{P}$}^{(q+1)}_p} \left\{\phi(\overline{\bfP}^{(q+1)}_1), \ldots , \phi(\overline{\bfP}^{(q+1)}_p)\right\}$$

\item \hspace{0.1cm} Set ${\bfP}^{(q+1)} := \overline{\bfP}^{(q+1)}_k$, where

$\left.{\bfP}^{(q+1)}= (P^{(q+1)}_1,...,P^{(q+1)}_p)\right.$

\item \hspace{0.1cm} {\bf If} $|\phi({\bfP}^{(q+1)}) - \phi({\bfP}^{(q)}) |\leq \epsilon $

\item \hspace{0.3cm} {\bf Stop}

\item \hspace{0.1cm} {\bf end}

\item {\bf end}

\vspace{0.20cm}

\end{enumerate}

\hrule

\vspace{0.2cm}

 Sequence $\{{\bfP}^{(q+1)}\}$ converges to a coordinate-wise minimum point of objective function $\phi({\bfP})$ which is a local minimum of (\ref{p1ph1p}).  Section \ref{convergence} provides more associated  details.

\begin{theorem}\label{th012}
For  $\bfP^{(q+1)}$ determined by Algorithm 1, the error associated with the proposed model of the second order WSN  is represented as
\begin{multline}\label{eq71}
\|{\bf x}-\bfP^{(q+1)}({\bf z})\|^2_\Omega = \|E_{xx}^{1/2}\|^2-\|E_{xz}(E_{zz}^{1/2})^\dagger\|^2  \\+\|E_{xz}(E_{zz}^{1/2})^\dagger-\bfP^{(q+1)}E_{zz}^{1/2}\|^2.
\end{multline}
\end{theorem}

\begin{IEEEproof} The proof follows from (\ref{xfy1p}).

\end{IEEEproof}
\begin{remark}\label{rem0}
Although (\ref{eq71}) represents a posteriori error, it is convenient to use during a testing phase.
\end{remark}
\begin{remark}\label{rem1}
For $p=1$, the formula in (\ref{whf1j}) coincides with the second degree transform (SDT) proposed in \cite{2001309}.  Therefore, the transform represented by $\displaystyle \p (\zz)=\sum_{j=1}^p \p_j(\zz_j)$  (see Definition \ref{def1}) where $P_j$ solves  $\min_{ P_j\in {\mathbb R}_{r_j}} \left\|H - \sum_{i=1}^p P_iG_i \right\|^2$, for $j=1,\ldots,p$,  can be regarded { the multi-compressor  SDT}. Further, Algorithm 1 represents the version of the MBI method that uses the multi-compressor  SDT. Therefore, the WSN model in the form  $\p^{(q+1)}(\zz)=\sum_{j=1}^p \p_j^{(q+1)}(\zz_j)$ where $P^{(q+1)}_1,\ldots,P^{(q+1)}_p$ are determined by Algorithm 1 can be interpreted as a transform as well. We call this transform   the multi-compressor  SDT-MBI.
\end{remark}

\subsection{Determination of Initial Iterations }\label{initial_iterations}

We determine initial iterations $P_j^{(0)}$, for $j=1,\ldots,p$, for Algorithm 1 as follows.
Let us denote $\x = [\x^T_1,\ldots,\x_p^T]^T$ where  $\x_i \in L^{2}(\Omega,{\mathbb R}^{m_i})$,  $i=1,\ldots,p$ and $m_1 + \ldots +m_p = m$. Suppose that matrix $T\in \rt^{m\times r}$ is given by
$
 T=\diag (T_{11},\ldots, T_{pp})
$
where $T_{jj}\in\rt^{m_j\times r_j}$, for $j=1,\ldots,p$.  Then
\begin{eqnarray}\label{b}
& & \left\|\left[ \begin{array}{c}
\x_1 \\
\vdots\\
\x_p\end{array} \right] - \diag (\ttt_{11},\ldots, \ttt_{pp}) \left[ \begin{array}{c}
\sss_1 (\zz_1) \\
\vdots\\
\sss_p(\zz_p)\end{array} \right] \right\|^2_{\Omega} \nonumber\\
& = & \left\|\left[ \begin{array}{c}
\x_1 - \p_1(\zz_1)\\
\vdots\\
\x_p - \p_p(\zz_p)\end{array} \right]  \right\|^2_{\Omega}\nonumber\\
& = & \sum_{j=1}^p \|\x_j - \p_j (\zz_j)\|^2_{\Omega}
\end{eqnarray}
and
\begin{eqnarray*}
\hspace{-0.5cm} \min_{\mathcal{P}_1\in \mathcal{R}_{r_1},\ldots, \mathcal{P}_p\in  \mathcal{R}_{r_p}}\sum_{j=1}^p \|\x_j - \p_j (\zz_j)\|^2_{\Omega}
 \hspace{-0.1cm} =  \hspace{-0.1cm} \sum_{j=1}^p \min_{P_j\in \mathcal{R}_{r_j}} \|\x_j - \p_j (\zz_j)\|^2_{\Omega}.
 \end{eqnarray*}
As a result, in this case, problem (\ref{p1p}) is reduced to the  problem of finding $\p_j$ that solves
 \begin{eqnarray}\label{f1pp1}
 \min_{\mathcal{P}_j\in \mathcal{R}_{r_j}} \left\|\x_j -  \mathcal{P}_j (\zz_j)\right\|^2_{\Omega},
  \end{eqnarray}
for $j=1,\ldots,p$. Its solution is given by \cite{2001309}
\begin{equation}\label{fj02}
\widehat{P}_j=\left[E_{x_j z_j}(E^\dagger_{z_jz_j})^{1/2}\right]_{r_j}(E^\dagger_{z_jz_j})^{1/2} (I + \widetilde{K}j)
\end{equation}
where $\widetilde{K}j = \widetilde{M}_j\left(I-L_{E_{z_jz_j}^{1/2}}\right)$ and $\widetilde{M}_j$ is an arbitrary matrix. Then the initial iterations for Algorithm 1 are defined by
 \begin{eqnarray}\label{fj0xyj}
P_j^{(0)} = \widehat{P}_j,\quad \text{for $j=1,\ldots,p$}.
\end{eqnarray}

\subsection{ Models of Sensors and Fusion Center}\label{models}

To determine the sensor model $B_j$ in the form (\ref{qqqj})--(\ref{sssj02}), for $j=1,\ldots,p$, and the fusion center model $T =[T_1,\ldots, T_p]$ we need to determine  $S_{j, 0}$ $S_{j, 1}$, $S_{j, 2}$ such that $S_j = [S_{j, 0}, S_{j, 1}, S_{j, 2}]$ and $T_j$, for $j=1,\ldots,p$.   Matrices $S_{j, 0}$, $S_{j, 1}$, $S_{j, 2}$ and $T_j$ are evaluated by  $S^{(q+1)}_{j, 0}\in\rt^{r_j}$, $S^{(q+1)}_{j, 1}\in\rt^{r_j\times (2n_j+1)}$, $S^{(q+1)}_{j, 2}\in\rt^{r_j\times (2n_j+1)}$ and ${T}^{(q+1)}_j\in \rt^{m\times r_j}$, respectively, as follows. By Algorithm 1,  the mathematical model of the second degree  WSN is given by
\begin{eqnarray}\label{wpqtsj}
{\bfP}^{(q+1)}({\bf z}) & = & \sum_{j=1}^p P^{(q+1)}_j({\bf z}_j)\nonumber\\
                        & = & \sum_{j=1}^p {T}^{(q+1)}_j {S}^{(q+1)}_j({\bf z}_j)\nonumber\\
                        & = & {T}^{(q+1)}\left(
                                          \begin{array}{c}
                                            {S}_1^{(q+1)}({\bf z}_1) \\
                                            \vdots \\
                                            {S}_p^{(q+1)}({\bf z}_p) \\
                                          \end{array}
                                        \right),
\end{eqnarray}
where ${\bfP}^{(q+1)} = (P^{(q+1)}_1,...,P^{(q+1)}_p)$  and $P^{(q+1)}_j={T}^{(q+1)}_j{S}^{(q+1)}_j$. By step 7 of Algorithm 1,
$$P^{(q+1)}_j= \widehat{P}^{(q+1)}_j= \left[Q^{(q)}_jR_{G_j}\right]_{r_j}G_j^\dagger (I+K_j)\quad$$ or $P^{(q+1)}_j={P}^{(0)}_j,$
where ${P}^{(0)}_j$ is represented by (\ref{fj02})-(\ref{fj0xyj}).
For the case when  $P^{(q+1)}_j = \widehat{P}^{(q+1)}_j$, we write
\begin{equation}\label{svdqrj}
\displaystyle U_{r_j}\Sigma_{r_j}V_{r_j}^T = \left[Q^{(q)}_jR_{G_j}\right]_{r_j},
 \end{equation}
 where  $\displaystyle U_{r_j}\Sigma_{r_j}V_{r_j}^T$ is the truncated SVD taken with first $r_j$ singular values, $U_{r_j}\in \rt^{m\times r_j}$, $\displaystyle \Sigma_{r_j}\in \rt^{r_j\times r_j}$ and $V_{r_j}^T\in \rt^{r_j\times (2n+p)}$. Then
\begin{equation}\label{tq1j1}
{T}^{(q+1)}_j = U_{r_j} \quad \text{or}\quad {T}^{(q+1)}_j = U_{r_j}\Sigma_{r_j}
\end{equation}
where ${T}^{(q+1)}_j \in \rt^{m\times r_j}$ and ${T}^{(q+1)}_j \in \rt^{m\times r_j}$, and
\begin{equation}\label{sq1j1}
{S}^{(q+1)}_j  =  \Sigma_{r_j}V_{r_j}^T G_j^\dagger (I+K_j) \;\;\; \text{or} \;\;\; {S}^{(q+1)}_j  =  V_{r_j}^T G_j^\dagger (I+K_j),
\end{equation}
respectively, where ${S}^{(q+1)}_j\in \rt^{r_j\times (2n_j+1)}$. Here, $S^{(q+1)}_j = [S^{(q+1)}_{j, 0}, S^{(q+1)}_{j, 1}, S^{(q+1)}_{j, 2}]$.

As a result, for the $j$th sensor model represented by (\ref{qqqj})--(\ref{sssj02}), $S_{j, 0}$ $S_{j, 1}$ and $S_{j, 2}$ are given by $S^{(q+1)}_{j, 0}$, $S^{(q+1)}_{j, 1}$ and $S^{(q+1)}_{j, 2}$, respectively, which are determined from (\ref{sq1j1}) as blocks of matrix ${S}^{(q+1)}_j$.
The fusion center  model is given by  ${T}^{(q+1)} = [{T}^{(q+1)}_1,\ldots, {T}^{(q+1)}_p]$ where ${T}^{(q+1)}_j$, for $j=1,\ldots,p$, is determined by (\ref{tq1j1}).

For the case  when  $P^{(q+1)}_j = {P}^{(0)}_j$, where ${P}^{(0)}_j$ is represented by (\ref{fj02})-(\ref{fj0xyj}), matrices $T_j$, $S_{j, 0}$ $S_{j, 1}$ and $S_{j, 2}$ are evaluated
similar to matrices ${T}^{(q+1)}_j$, $S^{(q+1)}_{j, 0},$ $S^{(q+1)}_{j, 1}$ and $S^{(q+1)}_{j, 2}$ in (\ref{tq1j1}) and  (\ref{sq1j1}), respectively. In this case, in (\ref{svdqrj}),   $\left[Q^{(q)}_jR_{G_j}\right]_{r_j}$ should be replaced with $\left[E_{x_j z_j}(E^\dagger_{z_jz_j})^{1/2}\right]_{r_j}$ and in (\ref{sq1j1}), $G_j^\dagger (I+K_j)$ should be replaced with $(E^\dagger_{z_jz_j})^{1/2} (I + \widetilde{K}j)$.

\begin{figure}
\centering
\begin{tabular}{c@{\hspace*{5mm}}c}
\includegraphics[scale=0.5]{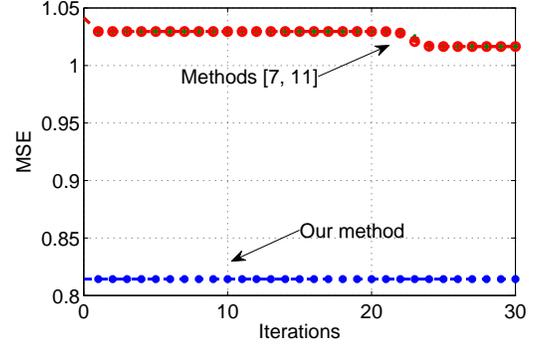}\\
(a) Example 1\\
\\
\includegraphics[scale=0.5]{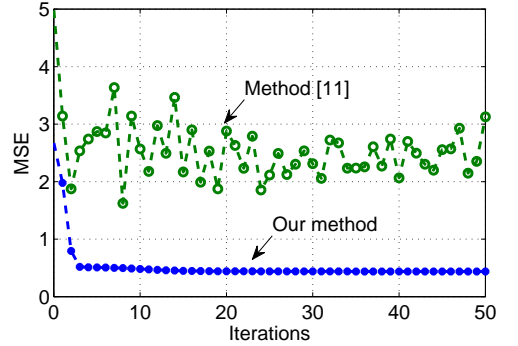}\\
(b) Example 2\\
\\
\includegraphics[scale=0.5]{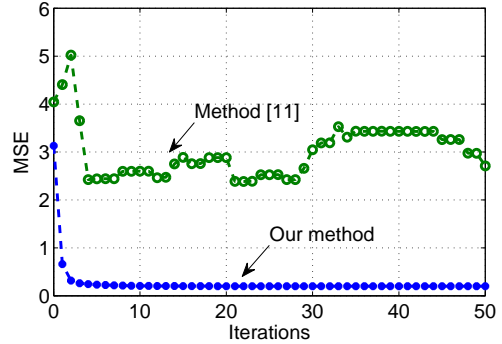}
(c) Example 2\\
\end{tabular}
  \caption{Diagrams of MSEs associated with our method and methods \cite{4276987,4475387}. In Fig. (a),  MSEs associated with methods \cite{4276987,4475387} coincide. }
  \label{ex:fig1}
\end{figure}

\section{Numerical Results and Comparisons}\label{sim}

We wish to illustrate the advantages of the proposed methodology by numerical examples where a comparison with  known methods  \cite{4276987,Song20052131, 1420805, 4016296,4475387,ma3098} is provided. The method \cite{4475387} represents a generalization of methods \cite{Song20052131, 1420805, 4016296} and therefore, we provide a numerical comparison with method \cite{4475387} which includes, in fact, a comparison with  methods  \cite{Song20052131, 1420805, 4016296} as well. Further, covariance matrices used in the  simulations associated with  Figs. \ref{ex:fig1} (b) and (c)   are singular, and therefore, method \cite{4276987} is not applicable (some more associated observations have been given at the end of  Section \ref{known}). Therefore, in  Figs. \ref{ex:fig1} (b) and (c),  results related to algorithm in \cite{4276987} are not given.

For the same reason, the method presented in \cite{ma3098} is not applicable as well. Moreover, the method in \cite{ma3098} is restricted to the case when the covariance matrix formed by the noise vector is block diagonal. This is not the case here.

In the examples below,  different types of noisy observed signals  and different compression ratios are considered. In all  examples, our method provides the better associated accuracy than that for the methods in \cite{4276987,4475387} (and methods in \cite{Song20052131, 1420805, 4016296} as well, because they follow from  \cite{4475387}).

\begin{figure*}[t!]
\centering
\begin{tabular}{ccc}
\includegraphics[scale=0.35]{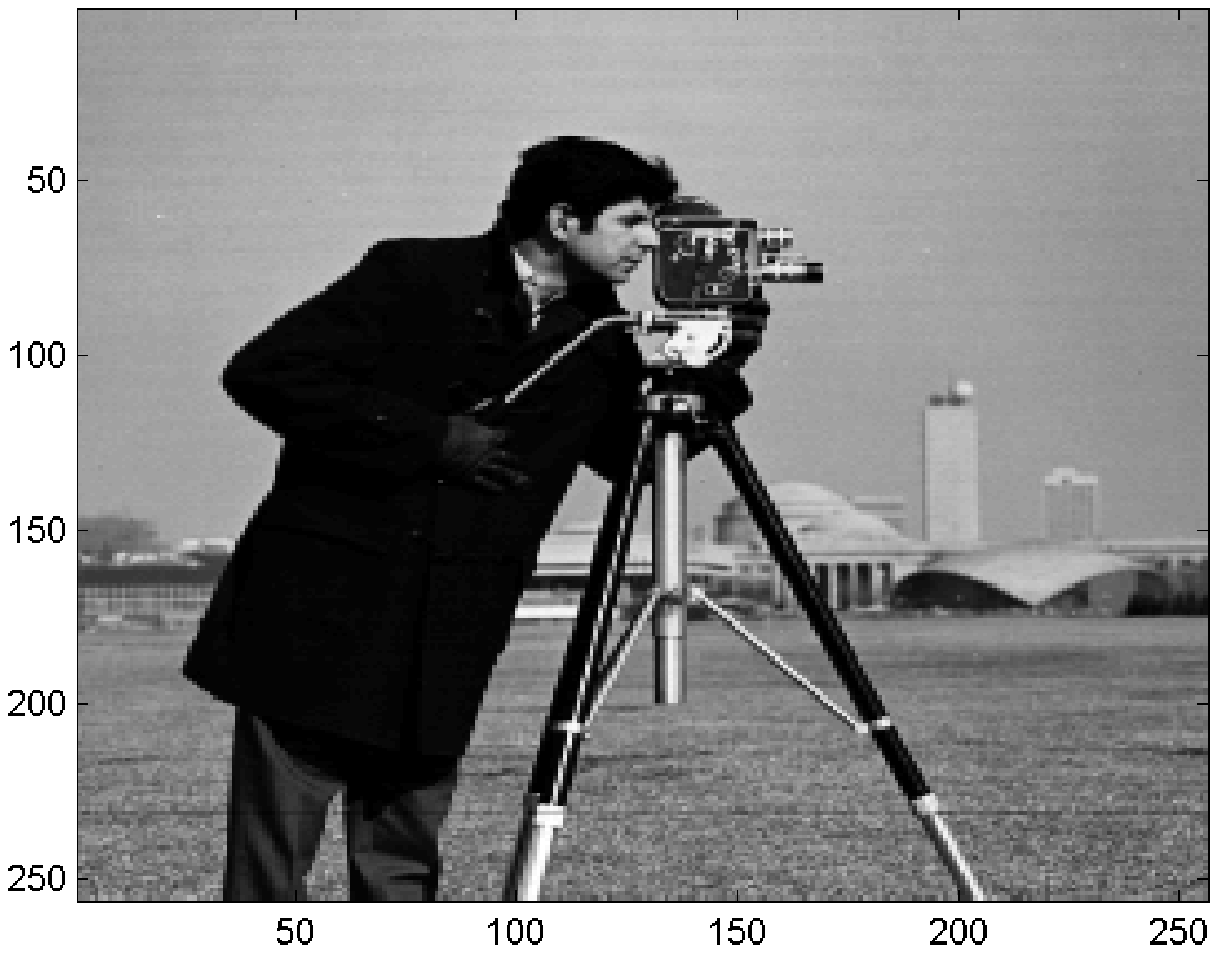} & \includegraphics[scale=0.35]{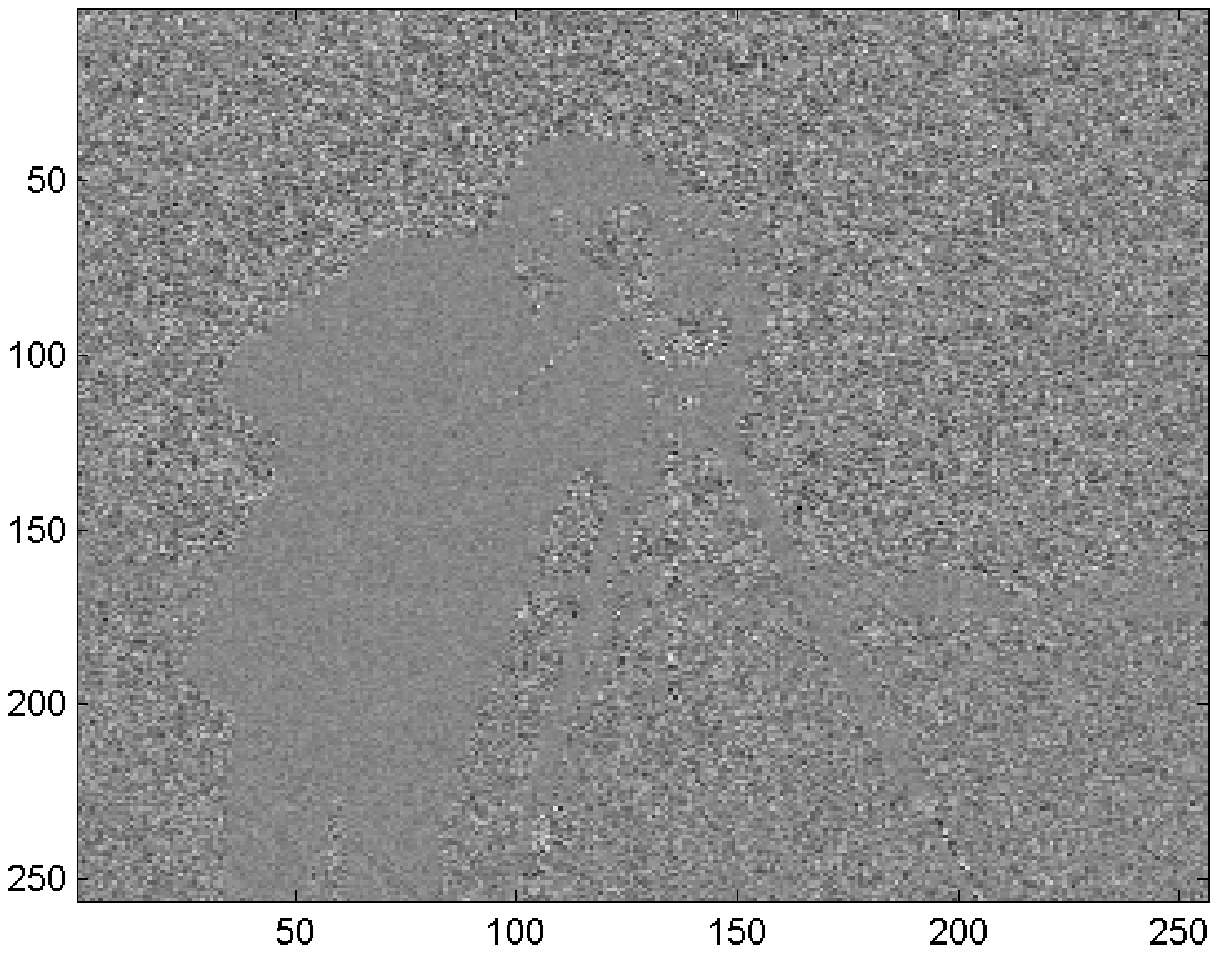} & \includegraphics[scale=0.35]{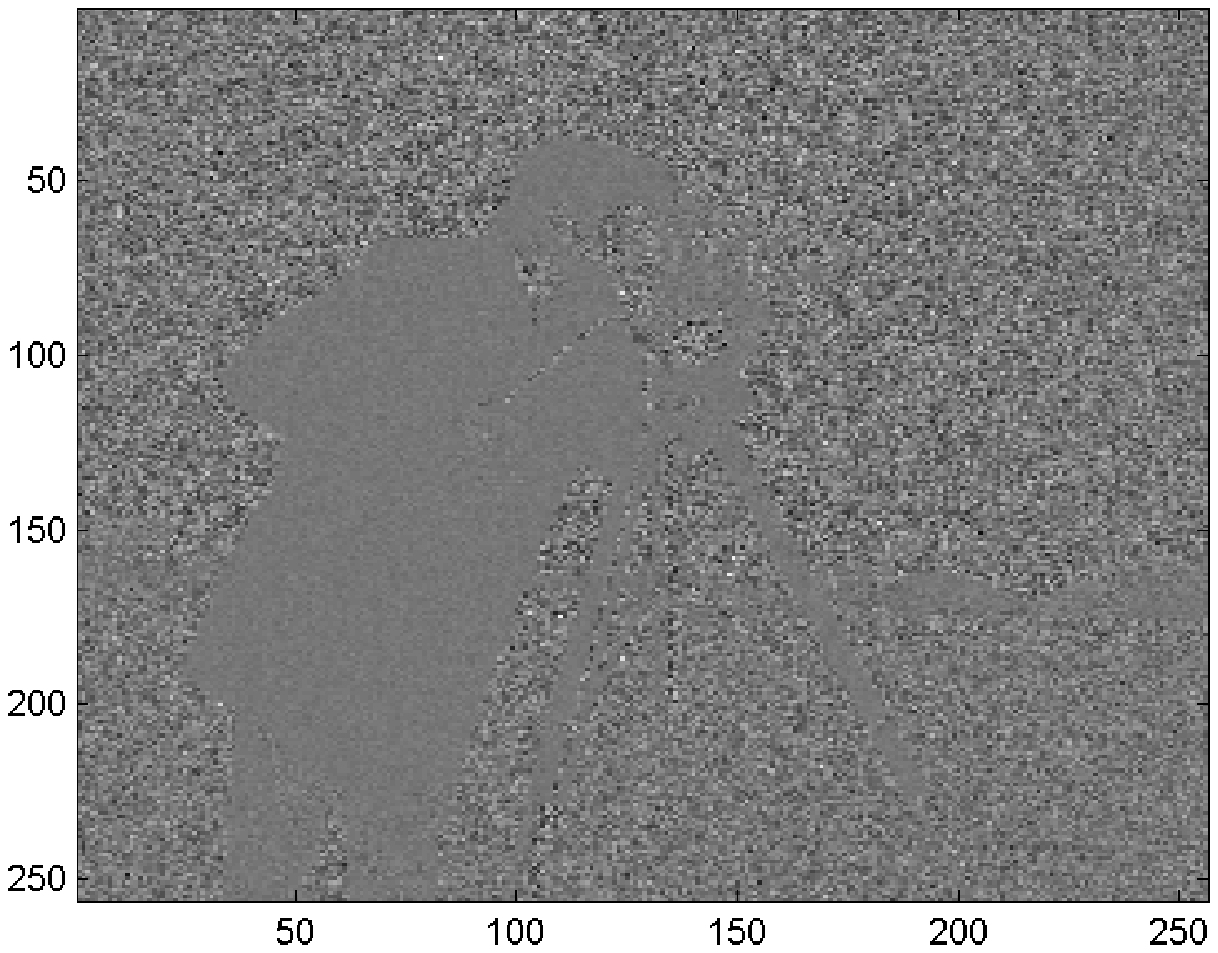}\\
(a) Original signal & (b) Data observed by sensor 1 & (c) Data observed by sensor 2\\
\\
\includegraphics[scale=0.35]{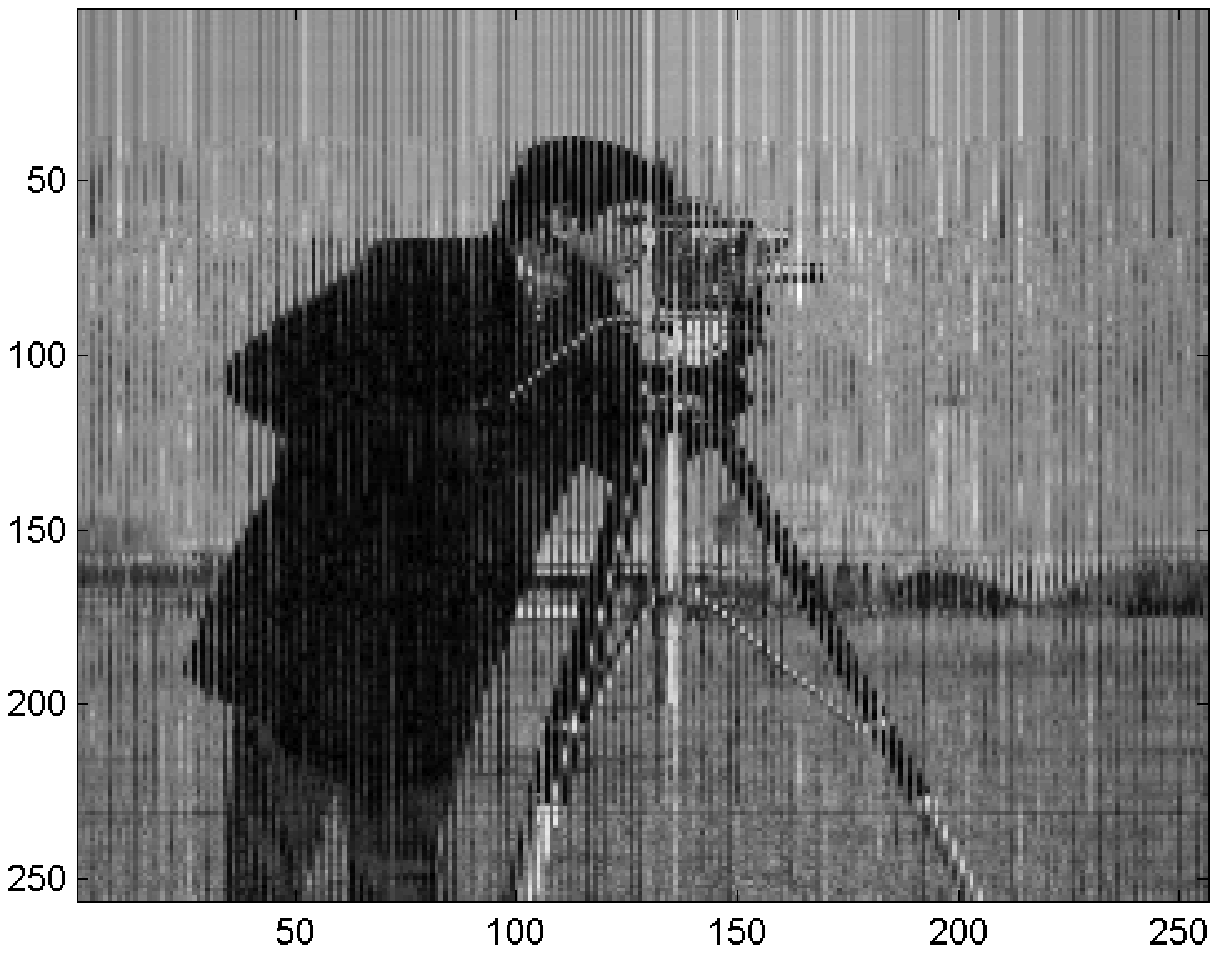} & \includegraphics[scale=0.35]{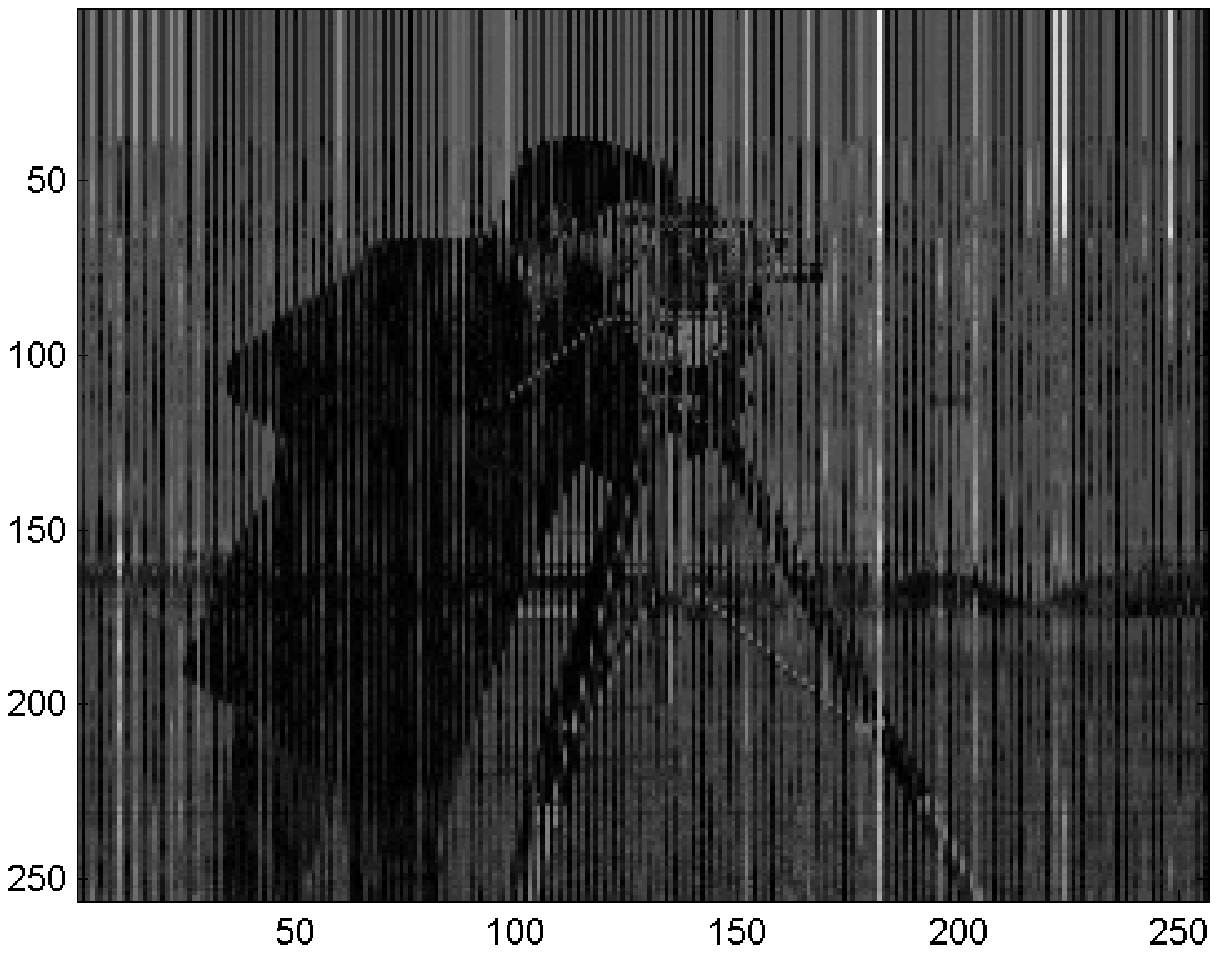} & \includegraphics[scale=0.35]{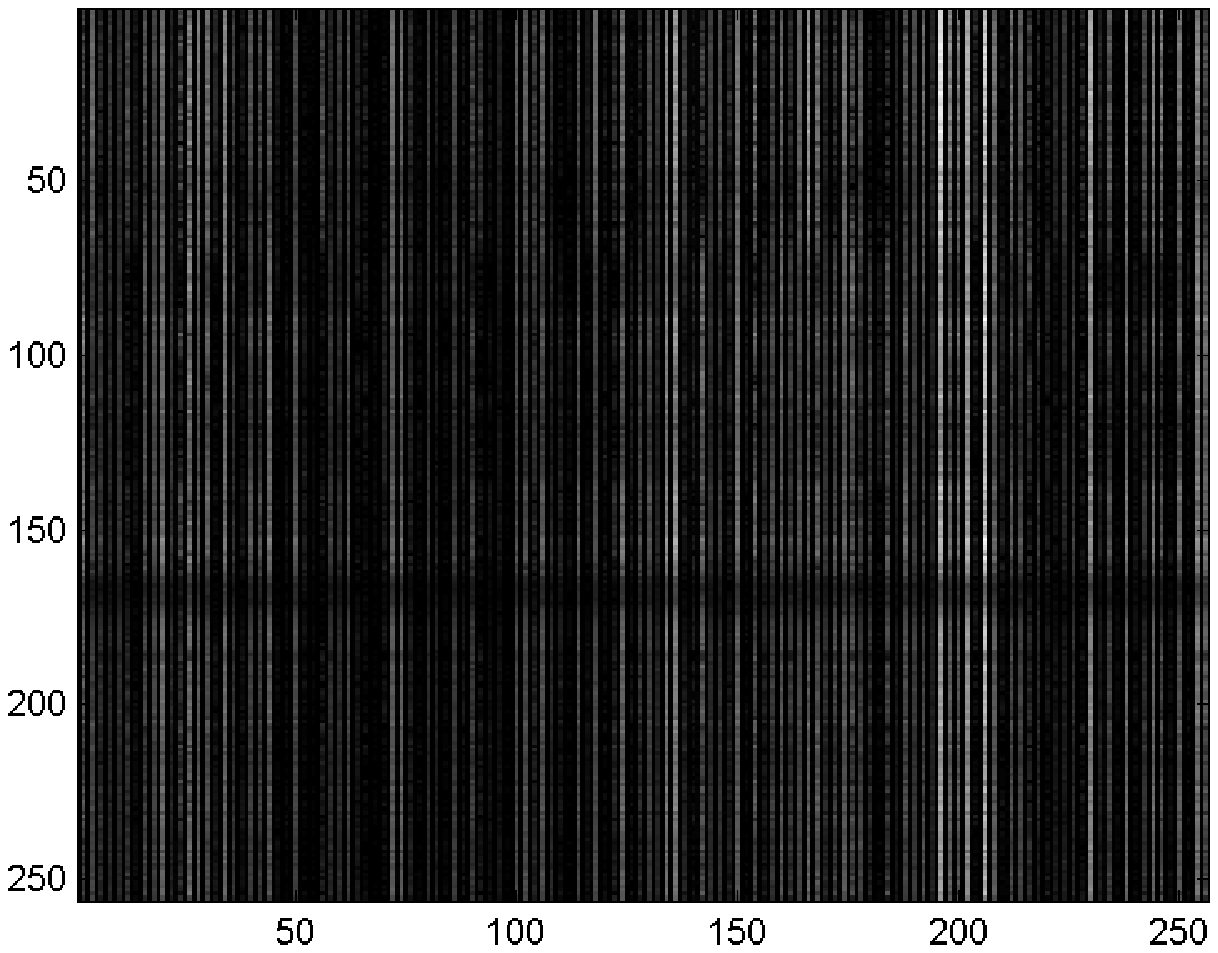}\\
(d) Signal estimate by our method & (e) Signal estimate by  method \cite{4475387} & (f) Signal estimate by  method \cite{4276987}\\
\end{tabular}
  \caption{Illustration to Example \ref{ex4}. }
  \label{ex:fig2}
\end{figure*}

\begin{example}\label{ex1}
 Suppose that the source signal $\x\in L^2(\Omega,\mathbb{R}^3)$ is a Gaussian random vector with mean zero and covariance matrix $E_{xx}= \left[
          \begin{array}{ccc}
           1 & 0.64 & 0.08 \\
           0.64 & 1 & 0.08 \\
           0.08 & 0.08 & 1 \\
         \end{array}
       \right].$
The observed vectors $\y_1$ and $\y_2$ are noisy version of $\x$, i.e., $\y_1=\x+\bxi_1$ and $\y_2=\x+\bxi_2$, where $\bxi_1$ and  $\bxi_2$  are Gaussian random vectors with mean zero and covariance matrices $E_{\xi_1, \xi_1} = \sigma_1^2 I$ and $E_{\xi_2, \xi_2} = \sigma_2^2 I$,  for $j=1,2$. The vectors $\x,$ $ \bxi_1$ and $\bxi_2$ are assumed to be independent. Then $E_{xy}=[E_{xx}\;E_{xx}]$ and
$E_{yy}  = \left[
  \begin{array}{cc}
    E_{y_1y_1} & E_{y_1y_2} \\
    E_{y_2y_1} & E_{y_2y_2} \\
  \end{array}
\right] =
\left[
       \begin{array}{ccc}
         E_{xx}+E_{\xi_1\xi_1} & E_{xx} \\
         E_{xx} & E_{xx}+E_{\xi_2\xi_2} \\
       \end{array}
     \right].$
For the sake of simplicity, in (\ref{ptsjyj})--(\ref{pjzzj1}),  we choose $\sss_{1,0}=\sss_{2,0}=\oo$. Therefore, $\zz_j=\left[
          \begin{array}{c}
            {\y_j} \\
            {\y_j}^2 \\
          \end{array}
            \right] $, for$j=1,2$, and as before (see Definition \ref{def1}), $\zz = [\zz_1^T,\vspace{1mm} \zz_2^T]^T$, i.e. ${\bf z}
        =\left[
          \begin{array}{c}
            {\bf y}_1 \\
            {\bf y}_1^2 \\
            {\bf y}_2 \\
            {\bf y}^2_2 \\
          \end{array}
            \right]
        =\left[
          \begin{array}{c}
            {\bf x}+\xi_1 \\
            ({\bf x}+\xi_1)^2 \\
            {\bf x}+\xi_2 \\
            ({\bf x}+\xi_2)^2 \\
          \end{array}
            \right]
        =\left[
          \begin{array}{c}
            {\bf x}+\xi_1 \\
            {\bf x}^2+\xi_1^2 \\
            {\bf x}+\xi_2 \\
            {\bf x}^2+\xi_2^2 \\
          \end{array}
            \right],$
because $\x,$ $ \bxi_1$ and $\bxi_2$ are independent. Matrices $E_{xz}$ and $E_{zz}$ are evaluated as follows. For a random variable $a$ with the probability density function $f_a(u)=\frac{1}{\sqrt{2\pi}}\exp\left(-\frac{u^2}{2\sigma_a^2}\right)$,  associated covariances are given by $E_{a a} = \sigma^2$ and $E_{a^2 a^2}=2\sigma^4$. Here, $\sigma^2$ is variance. For random variables $a$ and $b$ with the joint probability density function
 $f_{a,b}(u,v)=\frac{1}{2\pi\sqrt{1-\rho_{a,b}^2}}\exp\left(-\frac{u^2-2\rho_{a,b}uv+v^2}{2(1-\rho_{a,b}^2)}\right)$,  associated covariances are given by $E_{a b} = \rho_{a,b}$,  $E_{a^2 b}= E_{a b^2}=0,$ $E_{a^2 b^2}=2\rho^2_{a,b}$ and $E_{a^2 a}= E_{a a^2}=0$, where $\rho_{a,b}$ is the correlation parameter. Then
$E_{x^2x^2}=\left[
             \begin{array}{ccc}
                2       &  0.8192  &  0.0128\\
                0.8192  &  2       &  0.0128\\
                0.0128  &  0.0128  &  2
             \end{array}
           \right],$ $E_{\xi_1^2\xi_1^2}=2\sigma_1^4{I}$, $E_{\xi_2^2\xi_2^2}=2\sigma_2^4{I},$ $E_{xz}  =  [E_{xy}\;E_{xy^2}]  =  [E_{xy_1}\;E_{xy_2}\;E_{xy_1^2}\;E_{xy_1^2}]  =  [E_{xx}\;E_{xx}\;{\oo}\;{\oo}]$
and $E_{zz}$ is given in (\ref{cov_ma}).
\begin{figure*}
\hrule
\begin{eqnarray}\label{cov_ma}
E_{zz}  & = &   \left[
               \begin{array}{cc}
                 E_{yy} & E_{yy^2} \\
                 E_{y^2y} & E_{y^2y^2} \\
               \end{array}
             \right]\nonumber\\
        & = & \left[
                \begin{array}{cccc}
                  E_{y_1y_1} & E_{y_1y_2} & E_{y_1y_1^2} & E_{y_1y_2^2} \\
                  E_{y_2y_1} & E_{y_2y_2} & E_{y_2y_1^2} & E_{y_2y_2^2} \\
                  E_{y_1^2y_1} & E_{y_1^2y_2} & E_{y_1^2y_1^2} & E_{y^2_1y_2^2} \\
                  E_{y_2^2y_1} & E_{y_2^2y_2} & E_{y_2^2y_1^2} & E_{y^2_2y_2^2} \\
                \end{array}
              \right]\\
        & = & \left[
                \begin{array}{cccc}
                  E_{xx}+E_{\xi_1\xi_1} & E_{xx} & {\oo} & {\oo} \\
                  E_{xx} & E_{xx}+E_{\xi_2\xi_2} & {\oo} & {\oo} \\
                  {\oo} & {\oo} & E_{x^2x^2}+E_{\xi_1^2\xi_1^2} & E_{x^2x^2} \\
                  {\oo} & {\oo} & E_{x^2x^2} & E_{x^2x^2}+E_{\xi_2^2\xi_2^2} \\
                \end{array}
              \right]\nonumber.
\end{eqnarray}
\hrule
\end{figure*}
On this basis, Algorithm 1 produces the associated errors that were evaluated by (\ref{eq71}), for different $\sigma_j$ and $r_j$ with $j=1,2$. The errors  associated with  methods \cite{4276987,4475387} were estimated in a similar way. A typical example of the errors, for $\sigma_1=0.9$, $\sigma_2=0.65$ and $r_1=r_2=1,$ is represented in  Fig. \ref{ex:fig1} (a).
\end{example}

 \begin{example}\label{ex2}
 Here, it is supposed that covariance matrices  $E_{x x}$, $E_{x y}$ and  $E_{yy}$ are known from  a training phase.  We assume that, for $j=1,\ldots,p$, noisy observations are such that
$ 
\y_j = \aaa_j \x + \beta_j\bxi_j,
$ 
where $\aaa_j: L^2(\Omega,\mathbb{R}^{m}) \rightarrow L^2(\Omega,\mathbb{R}^{m})$ is a linear operator defined by  matrix $A_j\in \rt^{m\times m}$ with uniformly distributed random entries, $\x$ is a random vector with uniform distribution,  $\beta_j\in\rt$ and $\bxi_j$ is  noise represented by a Gaussian random vector with mean zero.


The errors associated with  the proposed method and the  method in \cite{4475387}, for the case of two and three sensors (i.e., for $p=2$ and $p=3$, respectively), and different choices of $m,n_j$ and $r_j$, for $j=1,2$ and $j=1,2,3$, are represented in Figs. \ref{ex:fig1} (b) and (c).
  Method \cite{4276987} is not applicable  since covariance matrices used in  these simulations    are singular.
Note that method \cite{4475387} is not numerically stable in these simulations. We believe this is because of the reason mentioned in Section \ref{known}.

 \end{example}

\begin{example}\label{ex4}
Here, we wish to illustrate the obtained  theoretical results in a more conspicuous and different way than that in the above examples.
To this end, we use training signals themselves, not only covariance matrices  as before. In this example, training reference signal $\x$  is simulated by its realizations, i.e. by  matrix ${\bf X}\in\rt^{m\times k} $ where each column  represents a realization of the signal.  To represent the obtained results in a visible way, signal ${\bf X}\in\rt^{m\times k}$ is chosen as the known image `Cameraman' given by the $256\times 256$ matrix -- see Fig. \ref{ex:fig2} (a), i.e. with $m, k =256$. Then $X\in\rt^{256\times 128} $.
A sample $X\in\rt^{m\times s} $ with $s < k$ is formed from $\bfX$ by choosing the even columns, i.e. $s=128$.

Further, we consider the WSN with two sensors, i.e. with $p=2$, where the observed signal $\bf Y_j$, for $j=1,2$, is simulated as
$
{\bf Y}_j = A_j * \bfX + \beta_j \up_j,
$
where $A_j\in\rt^{256\times 256} $ and $\up_j\in\rt^{256\times 256} $  have random entries, chosen from a normal distribution with mean zero and variance one, $A_j * X $ represents the Hadamard matrix product and $\beta_1=0.2$ and $\beta_2=0.1$.

 For $r_1= r_2=128$, the simulation results  are represented in Figs. \ref{ex:fig2} and \ref{ex:fig3}. Our method and methods from  \cite{4276987} and \cite{4475387} have been applied to the above signals with $50$ iterations each. The associated errors  are evaluated in the form $\|{\bf X}(i) - \widehat{\bf X} (i)\|_2^2$, for $i=1,\ldots,256$, where $\widehat{\bf X}$ is the reconstruction of $\bf X$ by  the  method we use (i.e. by our method or methods in \cite{4276987} and \cite{4475387}), and ${\bf X}(i)$ and  $\widehat{\bf X} (i)$ are $i$th columns of matrices ${\bf X}$ and $\widehat{\bf X}$, respectively.

 Method  \cite{4276987} has only been developed in terms of the inverses of certain covariance matrices. In this example, the covariance matrices are singular.  Consequently, the estimate by method \cite{4276987} has the form represented in  Fig. \ref{ex:fig2} (f).
Due to singularity of covariance matrices, the error associated with method \cite{4276987} is very large. Therefore, those results are not included in Fig. \ref{ex:fig3}.

Method  \cite{4276987} was applied in terms of pseudo-inverse matrices as it suggested in the section ``Appendix I'' of \cite{4276987}. Nevertheless, the associated estimate (see Fig. \ref{ex:fig2} (e)) is still not so accurate as that by the proposed method. As mentioned in Example \ref{ex2}, it might be, in particular, because  of the reason considered in Section \ref{known}.

 Similar to the other examples, Figs. \ref{ex:fig2} and \ref{ex:fig3} demonstrate  a more accurate   signal reconstruction associated  with the proposed method  than those associated with previous methods.

\end{example}

\begin{figure}[h!]
  \centering
  \includegraphics[height=5.5cm,width=6cm]{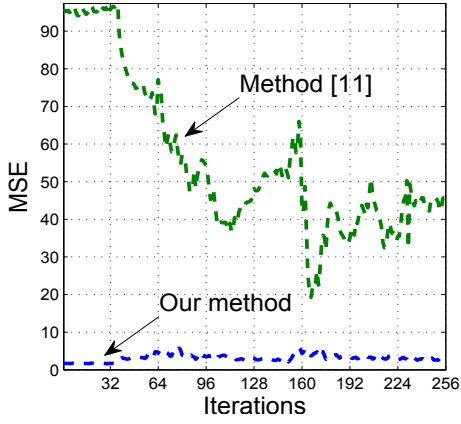}
 \caption{Example \ref{ex4}: Diagrams of MSEs associated with our method and method \cite{4475387}. MSE associated with method \cite{4276987} is very large and, therefore, it is not given here.} \label{ex:fig3}
\end{figure}

\section{Second Degree WSN with Nonideal Channels}\label{nonideal}

In the above sections, we dealt with the WSN model with ideal links between the sensors and the fusion center. Here, we consider nonideal channels which comprise multiplicative signal fadding and additive noise. These effects are modeled by operators  $\n_j : L^2(\Omega,\mathbb{R}^{r_j}) \rightarrow L^2(\Omega,\mathbb{R}^{r_j})$, for $j=1,\ldots,p$, which are depicted in Fig. \ref{fig1} (a). Each $\n_j$ is such that $\n_j(\uu_j) = \df_j(\uu_j) + \eet_j$ where the channel linear operator $\df_j$  and $\eet\in L^2(\Omega, \rt^{r_j})$ is random noise { with zero mean} which is uncorrelated with $\x$, $\y_j$, $\df_j$ and across the channels, i.e., $E_{\eta_i \eta_j} = \oo$ for $i\neq j$. Operator $\df_j$ is represented by matrix $D_j\in \rt^{r_j\times r_j}$ similar, in particular, to $\bb_j$ in (\ref{eq3}).
  Matrices $D_j$ and $E_{\eta_j \eta_j}$, for $j=1,\ldots,p$, are available at the fusion center. As in \cite{4276987} it is assumed that channel matrix $D_j$ and matrices $E_{\eta_j \eta_j}$ can be acquired similar to covariances $E_{xy}$ and $E_{yy}$. At the same time,  unlike \cite{4276987} here, it is not required that  $D_j$ and $E_{\eta_j \eta_j}$ are necessarily invertible.

In the case of nonideal channels, we deal with the problem as follows. Find models of the sensors $\mathcal{B}_1,\ldots,\mathcal{B}_p$ and the fusion center $\ttt = [\mathcal{T}_1,\ldots, \mathcal{T}_p]$ that solve
  \begin{eqnarray}\label{nb1bpd1dp}
  \min_{\substack{\mathcal{T}_1,\ldots, \mathcal{T}_p, \\ \mathcal{B}_1,\ldots,\mathcal{B}_p}}\left\|{\bf x}-\sum_{j=1}^p \ttt_j \left [\df_j\mathcal{B}_j({\bf y}_j)+\eet_j\right]\right\|_\Omega^2.
\end{eqnarray}
On the basis of (\ref{qqqj2}), the above can equivalently be represented as the problem of finding $\mathcal{S}_1,\ldots,\mathcal{S}_p$ and $\mathcal{T}_1,\ldots, \mathcal{T}_p$ that solve
  \begin{eqnarray}\label{ns1spd1dp}
  \min_{\substack{\mathcal{T}_1,\ldots, \mathcal{T}_p, \\ \mathcal{S}_1,\ldots,\mathcal{S}_p}}\left\|{\bf x}-\sum_{j=1}^p \ttt_j \left [\df_j\mathcal{S}_j({\bf z}_j)+\eet_j\right]\right\|_\Omega^2.
\end{eqnarray}

\subsection{Greedy Method for  Solution of Problem $(\ref{ns1spd1dp})$}\label{greedy ni}

An exact solution to problem (\ref{ns1spd1dp}) is unknown. By this reason, we apply a greedy approach to (\ref{ns1spd1dp}) in the form of an
 iterative  procedure.  At each  iteration,   the idea of the MBI method is combined with the optimal solutions to two pertinent minimization problems, so that one solution determines the optimal model of the sensor and another solution determines the optimal model of the fusion center. By this reason, it is called the alternating iterative (AI) procedure.

First, in Section \ref{prelim1}, we provide the solutions to the minimization problems used in the AI procedure. In Section \ref{descr1}, a detailed description of the steps used at each iteration of the AI procedure is provided. Then the AI procedure itself is considered in Section \ref{algorithm2}.

\subsubsection{Preliminaries for AI Procedure}\label{prelim1}


Let us denote  $$\psi (\bfP)= \displaystyle\left\|{\bf x}-\sum_{j=1}^p \ttt_j \left [\df_j\mathcal{S}_j({\bf z}_j)+\eet_j\right]\right\|_\Omega^2$$ where
$\bfP = (T_1,\ldots, T_p,$ $S_1,\ldots, S_p)$.

\begin{theorem}\label{th1n}
Let  $\w_j =\df_j\mathcal{S}_j({\bf z}_j)+\eet_j$, for $j=1,\ldots,p$, and  $\w = [\w^T_1,\ldots,\w^T_p]^T$. Let $K = M(I - L_{E_{w  w}})$ where $M$ be an arbitrary matrix. For given $S_1,\ldots, S_p$, optimal matrix $T$ minimizing $\psi (\bfP)$ is given by
\begin{eqnarray}\label{t1exw}
T = E_{x  w} E_{w w}^\dag (I+ K).
\end{eqnarray}
The associated error is represented by
\begin{eqnarray}\label{minbpq1}
\min_T \psi ({\bfP}) = \|E_{xx}^{1/2}\|^2 - \|E_{xw}(E_{ww}^{1/2})^\dag\|^2.
\end{eqnarray}

\end{theorem}

\begin{IEEEproof} It follows that  $\psi (\bfP)$ can be represented as
\begin{eqnarray}\label{bfxtwj11}
\psi (\bfP)=  \left\|{\bf x}-T\w\right\|_\Omega^2.
\end{eqnarray}
Then (\ref{t1exw}) and (\ref{minbpq1}) follow from  \cite{tor5277}. If  given matrices $S_1,\ldots, S_p$ are optimal in the sense of minimizing $\psi (\bfP)$  then Theorem  \ref{th1n} returns the globally optimal solution of (\ref{ns1spd1dp}).

\end{IEEEproof}

\begin{theorem}\label{th2n}
Let ${\x}_{(j)} = \x - \displaystyle\sum_{i=1\atop i\neq j}^p \ttt_i  [\df_i\mathcal{S}_i({\bf z}_i)+\eet_i]$. For given $T$ and $\left\{ S_i\right\}_{i=1, i\neq j}^p$,  optimal matrix $S_j$ minimizing $\left\|{\x}_{(j)}-\ttt_j\df_j\sss_j(\zz_j)\right\|_\Omega^2$ is given by
\begin{eqnarray}\label{sjtjdjkj}
S_j = (T_j D_j)^\dag E_{x_{(j)} z_j} E_{z_j z_j}^\dag(I+ K_j),
\end{eqnarray}
where $K_j = M_j(I - L_{E_{z_j z_j}})$ and $M_j$ is arbitrary. The associated error is represented by
\begin{eqnarray}\label{minsj}
& & \hspace{-0.5cm} \min_{S_j} \left\|{\x}_{(j)}-\ttt_j\df_j\sss_j(\zz_j)\right\|_\Omega^2 \nonumber\\
& = & \left \|E_{x_{(j)}x_{(j)}}^{1/2}\right\|^2-\left\|E_{x_{(j)}z_j}(E_{z_jz_j}^\dagger)^{1/2}\right\|^2 \nonumber\\
&  & +\left\|E_{x_{(j)}z_j}E_{z_jz_j}^{1/2\dagger}\left[I-T_jD_j(T_j D_j)^\dag\right]\right\|^2.
\end{eqnarray}
\end{theorem}

\begin{IEEEproof}
We write
\begin{eqnarray}\label{bfxjsjp}
& & \hspace{-1.5cm} \left\|{\bf x}-\sum_{j=1}^p \ttt_j \left [\df_j\mathcal{S}_j({\bf z}_j)+\eet_j\right]\right\|_\Omega^2\nonumber\\
& = &\left\|{\x}_{(j)}-\ttt_j\df_j\sss_j(\zz_j)\right\|_\Omega^2\nonumber\\
 & = & \left \|E_{x_{(j)}x_{(j)}}^{1/2}\right\|^2-\left\|E_{x_{(j)}z_j}(E_{z_jz_j}^\dagger)^{1/2}\right\|^2\nonumber\\
 & & + \left\|E_{x_{(j)}z_j}E_{z_jz_j}^{1/2\dagger}-T_jD_jS_jE_{z_jz_j}^{1/2}\right\|^2.
\end{eqnarray}
By \cite{tor5277}, $S_j$ in (\ref{sjtjdjkj}) minimizes $\left\|E_{x_{(j)}z_j}E_{z_jz_j}^{1/2\dagger}-T_jD_jS_jE_{z_jz_j}^{1/2}\right\|^2$ and hence, minimizes  $$\left\|{\x}_{(j)}-\ttt_j\df_j\sss_j(\zz_j)\right\|_\Omega^2.$$ The error representation in  (\ref{minsj}) follows from (\ref{sjtjdjkj}) and (\ref{bfxjsjp}).

  Note that if  given matrices  $T$ and $\left\{ S_i\right\}_{i=1, i\neq j}^p$ are optimal in the sense of minimizing $\psi (\bfP)$  then Theorem  \ref{th2n} returns the globally optimal solution of (\ref{ns1spd1dp}).
\end{IEEEproof}

\subsubsection{Description of Steps in AI Procedure}\label{descr1}

Before describing the AI procedure in Section \ref{algorithm2}  that follows, here we describe the steps made at its each iteration. The steps are based on Theorems \ref{th1n} and \ref{th2n} provided above.

{\em $1$st step.} Given $T^{(0)}, S^{(0)}_1,\ldots, S^{(0)}_p$, compute
\begin{eqnarray}\label{t1exw0}
T^{(1)} = E_{x  w^{(0)}} E_{w^{(0)}  w^{(0)}}^\dag (I+ K^{(1)}).
\end{eqnarray}
Here, $E_{x  w^{(0)}}= [E_{x  w^{(0)}_1},$ $\ldots,$ $E_{x  w^{(0)}_p}]$ and $E_{w^{(0)}  w^{(0)}} = \{E_{w^{(0)}_i  w^{(0)}_j} \}_{i,j=1}^p$, where, for  $j=1,\ldots,p$,
\begin{eqnarray}\label{exw0jxzj}
E_{xw^{(0)}_j} & = & E_{xz_j}(S^{(0)}_j)^T D_j^T, \nonumber\\
E_{w^{(0)}_iw^{(0)}_j} & = & D_iS^{(0)}_iE_{z_iz_j}(S^{(0)}_j)^T D_j^T+E_{\eta_i\eta_j},\\
K^{(1)} & = & M^{(1)} (I - L_{E_{w^{(0)}  w^{(0)}}})\nonumber
\end{eqnarray}
where $M^{(1)} $ is an arbitrary matrix. Then $T^{(1)}_1,\ldots, T^{(1)}_p$ are determined as blocks of matrix $T^{(1)} = [T^{(1)}_1,\ldots, T^{(1)}_p]$ given by (\ref{t1exw0}).

{\em $2$nd step.}
Given $\{T^{(1)}_1\}_{j=1}^p$ and  $\{ S^{(0)}_i\}_{i=1, i\neq j}^p$, compute, for $j=1,\ldots,p$,
\begin{eqnarray}\label{s1jt1jdj}
\widehat{S}^{(1)}_j = (T^{(1)}_j D_j)^\dag E_{x_{(j,0)} z_j} E_{z_j z_j}^\dag(I+ K_j)
\end{eqnarray}
where
\begin{equation}
E_{x_{(j,0)} z_j}=E_{xz_j}-\sum_{i=1\atop i\neq j}^p T^{(1)}_iD_iS^{(0)}_i E_{z_iz_j},
\end{equation}
\begin{eqnarray}
K_j = M_j(I - L_{E_{z_j z_j}}),
\end{eqnarray}
where $M_j$ is arbitrary. Note that (\ref{t1exw0}) and (\ref{s1jt1jdj}) are evaluated on the basis of (\ref{t1exw}) and (\ref{sjtjdjkj}), respectively.

{\em $3$rd step.}
Denote
\begin{equation}
  \overline{\bfP}^{(1)}_j =  \left(T^{(1)}_1,...,T^{(1)}_p, S^{(0)}_1,...,S^{(0)}_{j-1},\widehat{S}_j^{(1)},S^{(0)}_{j+1},...,S^{(0)}_{p}\right),
\end{equation}
set ${S}_j^{(1)} := \widehat{S}_j^{(1)}$ and select $\overline{\bfP}^{(1)}_k$, for $k=1,\ldots,p$, such that  $\psi(\overline{\bfP}^{(1)}_k)$ is minimal among all $\psi (\overline{\bfP}^{(1)}_1),$ $\ldots,$ $\psi (\overline{\bfP}^{(1)}_k),$ $\ldots, \psi (\overline{\bfP}^{(1)}_p)$, i.e.,
\begin{equation}\label{}
  \overline{\bfP}^{(1)}_k =  \arg \min_{\mbox{\scriptsize\boldmath $\overline{P}$}^{(1)}_1,\ldots, \mbox{\scriptsize\boldmath $\overline{P}$}^{(1)}_p} \left \{\psi (\overline{\bfP}^{(1)}_1), \ldots, \psi (\overline{\bfP}^{(1)}_k),\ldots, \psi (\overline{\bfP}^{(1)}_p) \right\}.
\end{equation}
Then write
\begin{eqnarray}\label{bp1p1k}
{\bfP}^{(1)} := \overline{\bfP}^{(1)}_k
\end{eqnarray}
and denote ${\bfP}^{(1)} = (P^{(1)}_1,...,P^{(1)}_p)\in\rt_{r_1,\ldots,r_p}$. Then we repeat steps (\ref{t1exw0})--(\ref{bp1p1k}) with the replacement of $T^{(0)}, S^{(0)}_1,\ldots, S^{(0)}_p$  by $T^{(1)}, S^{(1)}_1,\ldots, S^{(1)}_p$ as follows. Compute
\begin{eqnarray*}
T^{(2)} = E_{x  w^{(1)}} E_{w^{(1)}  w^{(1)}}^\dag (I+ K^{(2)}),
\end{eqnarray*}
 where $ E_{x  w^{(1)}}$ and $E_{w^{(1)}  w^{(1)}}$ are evaluated similar to (\ref{exw0jxzj}) with subscript $(0)$  replaced by subscript $(1)$. Then similar to  (\ref{s1jt1jdj}) compute
 \begin{eqnarray}\label{s1jt1jdj2}
\widehat{S}^{(2)}_j = (T^{(2)}_j D_j)^\dag E_{x_{(j,1)} z_j} E_{z_j z_j}^\dag(I+ K_j)
\end{eqnarray}
where $E_{x_{(j,1)} z_j}=E_{xz_j}-\displaystyle\sum_{i=1\atop i\neq j}^p T^{(2)}_iD_iS^{(1)}_i E_{z_iz_j}$
and continue the computation with  other pertinent replacement in the superscripts as it is represented in Algorithm 2 that follows.

\subsubsection{Algorithm for AI Procedure}\label{algorithm2}
It follows from (\ref{bfxtwj11}) that the cost function can be written as
\begin{equation}\label{fbfperr}
\psi (\bfP) = \left\|E_{xx}^{1/2}\right\|^2 - \left\|E_{xw}(E_{ww}^{1/2})^\dag\right\|^2  + \left\|E_{xw}(E_{ww}^{1/2})^\dag - TE_{ww}^{1/2}\right\|^2.
\end{equation}

The process described above is summarized in Algorithm 2.
\\

\begin{figure}
\hrule

\vspace{0.1cm}

\noindent{\bf Algorithm 2:} Greedy approach to problem  (\ref{ns1spd1dp}): AI procedure

\vspace{0.1cm}

\hrule

\vspace{0.1cm}

\noindent{\bf Initialization:} $T^{(0)}, S_1^{(0)},...,S_p^{(0)}$, $D_1,...,D_p$ and $\epsilon>0$.

\vspace{0.25cm}

\hrule

\vspace{0.25cm}

\begin{enumerate}

\item {\bf for }$q=0,1,2,...$

\item \hspace{0.1cm} $E_{xw^{(q)}_j}=E_{xz_j}(S^{(q)}_j)^T D_j^T$

\item \hspace{0.1cm} $E_{w^{(q)}_iw^{(q)}_j}= D_iS^{(q)}_iE_{z_iz_j}(S^{(q)}_j)^T D_j^T+E_{\eta_i\eta_j}$

\item \hspace{0.1cm} $T^{(q+1)}=E_{xw^{(q)}}E_{w^{(q)}w^{(q)}}^\dagger(I+ K^{(q+1)})$

\item \hspace{0.1cm} $T^{(q+1)}_1=T^{(q+1)}(:,1:r_1)$

\item \hspace{0.1cm} {\bf for }$j=2,3,...,p$

\item \hspace{0.3cm} $k_j=\sum_{i=1}^j r_i$

\item \hspace{0.3cm} $T^{(q+1)}_j=T^{(q+1)}(:,r_{j-1}+1:k_j)$

\item \hspace{0.1cm} {\bf end}

\item \hspace{0.1cm} $\overline{\bf P}_0^{(q+1)}=(T^{(q+1)}_1,,...,T^{(q+1)}_p,S_1^{(q)},...S_p^{(q)})$

\item \hspace{0.1cm} {\bf for }$j=1,2,...,p$

\item \hspace{0.3cm} $E_{x_{(j, q)}z_j} = E_{xz_j}-\displaystyle\sum_{i=1\atop i\neq j}^p T^{(q+1)}_iD_iS^{(q)}_i E_{z_iz_j}$

\item \hspace{0.3cm} $\widehat{S}^{(q+1)}_j=(T^{(q+1)}_jD_j)^\dagger E_{x_{(j, q)}z_j}E_{z_jz_j}^\dagger (I + K_j),$

\item \hspace{0.3cm} $\overline{\bf P}^{(q+1)}_j=\left(T^{(q)},S_1^{(q)},...,S^{(q)}_{j-1},\widehat{S}^{(q+1)}_j,S_{j+1}^{(q)},...,,S_p^{(q)}\right)$

\item \hspace{0.3cm} $S_j^{(q+1)}=\widehat{S}_j^{(q+1)}$

\item \hspace{0.1cm} {\bf end}

\item \hspace{0.1cm} Choose $\overline{{\bfP}}^{(q+1)}_k$ such that

$\displaystyle\hspace{-0.5cm}\overline{{\bfP}}^{(q+1)}_k = \arg\min_{\overline{{\bfP}}^{(q+1)}_1,...,\overline{{\bfP}}^{(q+1)}_p} \left\{\psi (\bar{\bf P}^{(q+1)}_0), ...,  \psi (\bar{\bf P}^{(q+1)}_p)\right\}$

\item \hspace{0.1cm} Set ${\bfP}^{(q+1)}:=\overline{{\bfP}}^{(q+1)}_k$ where

${\bfP}^{(q+1)} = (T^{(q)},S_1^{(q)},...,S_p^{(q)})$

\item \hspace{0.1cm} {\bf If} $|\psi ({\bfP}^{(q+1)}) - \psi ({\bfP}^{(q)}) |\leq \epsilon$

\item \hspace{0.3cm} {\bf Stop}

\item \hspace{0.1cm} {\bf end}

\item {\bf end}

\end{enumerate}

\hrule

\end{figure}

On line 17, $\psi ({\bfP}^{(q+1)})$ is evaluated by (\ref{fbfperr}) with the replacement of $E_{xw_j}$, $E_{w_iw_j}$, $T$ with  $E_{xw^{(q)}_j}$, $E_{w^{(q)}_iw^{(q)}_j}$, $T^{(q+1)}$, respectively, which have been computed in lines 2 and 3.

Sequence $\{{\bfP}^{(q+1)}\}$ in Algorithm 2 converges to a coordinate-wise minimum point of $\psi({\bfP})$ which is a local minimum.  Section \ref{convergence2} provides more associated  details.

Similar to Algorithm 1, in a training stage, it is convenient to use the error representation  given in Theorem \ref{th012} where $P^{(q+1)}$ should be understood as that determined by Algorithm 2.

\subsection{Algorithm 2: Determination of   $T^{(0)}, S_1^{(0)},...,S_p^{(0)}$}\label{initial_iterations2}

To start  Algorithm 2, matrices $T^{(0)},S_1^{(0)},...,S_p^{(0)}$ should be known. They can be determined by the procedure detailed in Section \ref{models} for the case of the ideal channels. That is, for Algorithm 2, $T^{(0)}$ is determined as ${T}^{(q+1)} = [{T}^{(q+1)}_1,\ldots, {T}^{(q+1)}_p]$ where ${T}^{(q+1)}_j$, for $j=1,\ldots,p$, is given by (\ref{tq1j1}). For $j=1,\ldots,p$,  matrix $S_j^{(0)}$, for Algorithm 2, is determined as $S_j^{(q+1)}$ by (\ref{sq1j1}).

\subsection{ Models of Sensors and Fusion Center by Algorithm 2}\label{models2}

By (\ref{qqqj})--(\ref{sssj02}), the sensor model $B_j$ is defined by matrix $S_j= [S_{j, 0}, S_{j, 1}, S_{j, 2}]$. Matrix  $S_j$ is evaluated as $S_j^{(q+1)}$ at line 14.  Since $S_j^{(q+1)}$ can be written as $S_j^{(q+1)} = [S_{j, 0}^{(q+1)}, S_{j, 1}^{(q+1)}, S_{j, 2}^{(q+1)}]$ where $S_{j, k}^{(q+1)}$ is a block of $S_j^{(q+1)}$, for $k=0,1,2$, then $S_{j, 0}, S_{j, 1}, S_{j, 2}$ are evaluated as $S_{j, 0}^{(q+1)}, S_{j, 1}^{(q+1)}, S_{j, 2}^{(q+1)}$.

The fusion center $T = [T_1,\ldots, T_p]$ is evaluated as $T^{(q+1)} = [T_1^{(q+1)},\ldots, T_p^{(q+1)}]$ where $T_j^{(q+1)}$, for $j=1,\ldots,p$, is computed  on the lines from 3 to 7.

\section{Numerical Results and Comparison}\label{sim2}

Here, we illustrate the performance of the proposed approach with numerical modeling of the second degree WSN with two sensors and nonideal channels.

\begin{example}\label{ex41}
 Suppose that  source signal $\x\in L^2(\Omega,\mathbb{R}^4)$ is a Gaussian random vector with mean zero and covariance matrix
$$E_{xx}=\left[
  \begin{array}{cccc}
    1.000  &  0.580  &  0.275  &  0.450\\
    0.580  &  1.000  &  0.295  &  0.540\\
    0.275  &  0.295  &  1.000  &  0.215\\
    0.450  &  0.540  &  0.215  &  1.000
  \end{array}
\right].$$ Observed signals $\y_1$ and $\y_2$ are noisy versions of $\x$ such that $\y_j = \x + \delta^2_j \xi_j$, $\delta_j\in\rt$, for $j=1,2$, where $\xi_1$ and $\xi_2$ are Gaussian random vector with mean zero and covariance matrices $E_{\xi_i \xi_j} = \delta_j^2 I$, for $j=1,2$.
The vectors $\x$, $\xi_1$ and $\xi_2$ are independent. Channel matrices are $D_1=\left(\begin{array}{cc} 6 & 6 \\ 2 & 8 \\ \end{array} \right)$ and $D_2=\left(\begin{array}{cc} 0 & 5 \\ 0 & 5 \\ \end{array} \right)$. Note $D_2$ is singular, i.e., method in \cite{4276987} is not applicable here.
Suppose the channel noise $\eet_j$ is white, for $j=1,2$, i.e., $E_{\eta_j,\eta_j}=\gamma_j^2 I.$

The diagrams of the MSE, given in Fig. \ref{figcn11}, for $r_1=r_2=2$, $\delta_1=0.7$, $\delta_1=0.8$, $\gamma_1=0.6$,  $\gamma_2=0.5$, demonstrate the advantage of the second degree model over its particular case, the linear model.

\begin{figure}[h]
  \centering
  \includegraphics[width=8cm]{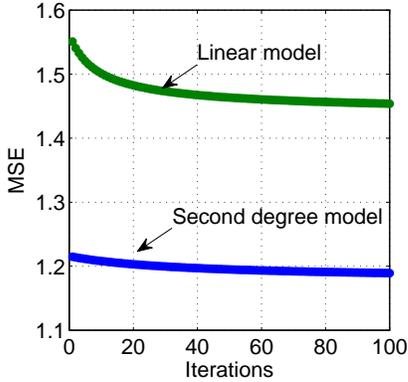}\\
 \caption{Diagrams of the MSE for the WSNs with nonideal channels.}
\label{figcn11}
\end{figure}

\end{example}

\section{Conclusion}
\label{conclusion}

We have addressed the problem of  multi-compression and recovery of  an unknown source stochastic signal  when
the signal cannot be observed centrally. In the mean square estimation framework, this means finding the optimal signal representation which minimizes the associated error. This scenario is  based on  multiple noisy signal observations made by distributed sensors.
Each sensor  compresses the observation and then transmits the compressed signal to
the fusion center that decompresses the signals in such a way that the original signal is
estimated within a prescribed accuracy.
By known techniques, models of the sensors and the fusion center have been developed from optimal {\em linear} transforms and therefore, the associated errors cannot be improved by any other linear transform with the same rank. At the same time, in some problems, the performance of the known techniques may not be as good as required.

In this paper, we  developed a new method for the determination of models of the sensors and the fusion center.  The contribution and advantages are as follows. First,  the proposed approach is based on the
 {\em non-linear} second degree transform (SDT) \cite{2001309}. The special condition has been established under which the proposed methodology may lead to the higher accuracy of signal estimation compared to known methods based on linear signal transforms.  Second, the SDT is combined with the  maximum block improvement  (MBI) method \cite{chen87,Zhening2015}}.  Unlike the commonly used block coordinate descent (BCD) method, the MBI method converges to a local  minimum under weaker conditions than those required by the BCD method (more details have been  provided in Sections \ref{known} and \ref{differences}). Third, the models of the sensors and the fusion center have been determined in terms of the pseudo-inverse matrices. Therefore, they are always well determined and numerically stable.  As a result, this approach mitigates to some extent the difficulties associated with the existing techniques. Since a `good' choice of the initial iteration gives  reduced
errors,   special methods for determining initial iterations have been considered.

The error analysis of the proposed method has been provided.


\section{Future Development}

The problem of  WSN modeling  with nonideal channels subject to a power constraint per sensor was tackled with different approaches considered, in particular, in \cite{4276987,Ribeiro1131,Xiao27,Junlin3746,Santos239}. It appears that this is a specialized topic and we intend to extend the  approach proposed in Section \ref{nonideal} to this important problem.
Another important topic relates to the analysis of the  influence   of the covariance estimation errors on models of the sensors and the fusion center. As mentioned before, special methods were developed, in particular,  in \cite{149980,Ledoit2004365,ledoit2012,Adamczak2009,Vershynin2012,won2013,schmeiser1991,yang1994}.  In our future research, we intend  to extend the known related results  to the WSN modeling with the proposed methodology.

\section{Acknowledgment}

The authors are grateful to anonymous reviewers for their useful suggestions that led to the significant improvement of the presented results.

\section{Appendix}\label{appendix}

\subsection{Advantages of Non-linear Approximation }\label{nonlinear}

 Advantages of non-linear approximation have been considered, in particular,  in \cite{Prenter341,Istr118,Bruno13,Akhiezer1992,Sandberg40,Mathews2001,Howlett353}   where the {\em non-linear} approximator is represented by the $q$-degree polynomial operator ${\mathbb P}_q : L^2\rightarrow L^2$  defined as
$ {\mathbb P}_q(\y) = {\aaa}_0 + {\aaa}_1(\y) + {\aaa}_2(\y,\y) +\ldots + {\aaa}_q(\underbrace{\y,\ldots,\y}_{{\scriptsize q }})$.
 Here, ${\aaa}_0$ stands for a constant operator and  ${\aaa}_k:(L^2)^k\rightarrow L^2$, for $k=1,\ldots,q$, is a $k$-linear operator.
 In broad terms, it is proven  that under certain conditions specified in  \cite{Prenter341,Istr118,Bruno13,Akhiezer1992,Sandberg40,Mathews2001,Howlett353}, a non-linear  transform $\it\Phi: L^2 \rightarrow L^2$, where $L^2:=L^2(\Omega,\rt^n)$, defined as $\x = \it\Phi(\y)$ can uniformly be approximated by $ {\mathbb P}_q(\y)$ to any degree of accuracy.  In other words, for any$\epsilon > 0$,  there is a $q$-degree polynomial $ {\mathbb P}_q(\y)$ such that, for an appropriate norm $\|\cdot\|$,
\begin{eqnarray}\label{xpyep}
\| \x - {\mathbb P}_q(\y)\| < \epsilon.
\end{eqnarray}
 For $q>1$, the $q$-degree polynomial $ {\mathbb P}_q(\y)$ has more parameters  to optimize compared to the linear case when ${\mathbb P}_1(\y) = {\aaa}_1(\y)$. That is why $ {\mathbb P}_q(\y)$ can provide any degree of accuracy.  Indeed, optimization of  $ {\mathbb P}_q(\y)$ is achieved by a variation of $q+1$ parameters, ${\aaa}_0$, ${\aaa}_1, \ldots,{\aaa}_q,$ while optimization of ${\mathbb P}_1(\y) = {\aaa}_1(\y)$ follows from choosing one parameter only, ${\aaa}_1.$

An optimal determination of  the $q$-degree polynomial $ {\mathbb P}_q(\y)$  is a hard research problem \cite{Deutsch2001}. Its solution  is subject to special conditions (convexity, for example) that are not satisfied for the problem in (\ref{f1p}) if $\f_j(\y_j)$ is replaced by  $ {\mathbb P}_q(\y)$.
For a particular case  in (\ref{f1p}), when $p=1$,  $q=1$ and ${\aaa}_0$ is the zero operator, this difficulty was surmounted  in \cite{4276987,Song20052131, 1420805, 4016296, 4475387,Schizas2007, dragotti2009, 5447742}  where a special approach for finding  $ {\mathbb P}_1(\y)$ that solves
\begin{equation}\label{p1r1}
\min_{ {\mathbb P}_1\in \mathcal{R}_{r_1}} \left\|{\bf x}- {\mathbb P}_1(\y)\right\|_\Omega^2.
\end{equation}
has been developed. In \cite{2001309}, for still  $p=1$ in (\ref{f1p}), a more general case has been considered, when $q=2$ and  ${\mathbb P}_2(\y)$ is represented in a special form given by $ {\mathbb P}_2(\y) = {\aaa}_1(\y) + {\aaa}_2(\y,\y)$, where ${\aaa}_2(\y,\y)= \widetilde{\aaa}_2(\y^2)$, $\widetilde{\aaa}_2$ is a linear operator and $\y^2$ is given by $\y^2(\omega) := ([\y_1(\omega)]^2,\ldots, [\y_n(\omega)]^2)^T$.  The solution of  the problem
\begin{eqnarray*}
&&\hspace{1cm}\min_{{\mathcal A}_1, {\widetilde{\mathcal A}}_2 } \left\|{\bf x}- {\mathbb P}_2(\y)\right\|_\Omega^2\\\
&&\mbox{s. t.} \quad \rank [{\aaa}_1, \hspace{1mm} \widetilde{\aaa}_2]\leq r\leq \min\{m, 2n\}\nonumber
\end{eqnarray*}
obtained in \cite{2001309} shows that  the increase in $q$ to $q=2$ allows us to achieve a better associated accuracy than that for the linear case, i.e. with  $q=1$, for the same compression ratio.

The result in \cite{2001309} has been  extended in  \cite{Torokhti20061431}, for  an arbitrary $q$ in ${\mathbb P}_q(\y)$ and still for $p=1$ in (\ref{f1p}). Due to a special structure of the polynomial ${\mathbb P}_q(\y)$ based on the  orthogonalisation procedure proposed in \cite{Torokhti20061431}, it has been shown in Theorem 2 in \cite{Torokhti20061431} that, for $p=1$ and $\f_1$ represented by  ${\mathbb P}_q(\y)$   in (\ref{f1p}),  the accuracy of $\x$ estimation in (\ref{f1p}) is further improved  when $q$ increases. The orthogonalisation procedure \cite{Torokhti20061431} cannot be used for the problem under consideration since the sensors should not communicate with each other. It is one of the reasons to develop the greedy approach represented in the above Sections \ref{differences}, \ref{structure}.

\subsection{Explanation for the Relation in  (\ref{eq3})}\label{Explanation}

In Fig. \ref{fig1} (a), the input of the $j$th sensor is  random vector ${\bf y}_j$ which is, by the definition,  a function from $\Omega$ to $\rt^{n_j}$. More specifically, to satisfy the boundedness condition,  $\y_j\in L^2(\Omega,\mathbb{R}^{n_j})$.
  Since a multiplication of {\em function} ${\bf y}_j$ by a {\em matrix} is not defined,  the sensor model is represented by linear operator $\bb_j$.  As a result, in the LHS of (\ref{eq3}), $\bb_j(\y_j)$ is a random vector  as well (i.e., a function from $\Omega$ to $\rt^{n_j}$) and then $[\bb_j(\y_j)](\omega)$, for any $\omega\in\Omega$, is a vector in $\mathbb{R}^{n_j}$.
   To define operator $\bb_j$, matrix $B_j$ is used. To be consistent,  operator $\bb_j$ and matrix $B_j$ relate to each other as in (\ref{eq3}).
In the RHS  of (\ref{eq3}),  ${\bf y}_j(\omega)$ is a vector in $\mathbb{R}^{n_j}$, for every $\omega\in\Omega$, and therefore, multiplication of  $B_j$ by the vector  ${\bf y}_j(\omega)$  is well defined.

\subsection{Justification of Algorithm 1}\label{convergence}

\subsubsection{Proof of Theorem \ref{th01}}
On the basis of (\ref{hpjgj1}),
 \begin{eqnarray*}
\min_{\mathcal{P}_j\in\mathcal{R}_{r_j} } \left\|{\bf x}-\sum_{j=1}^p \p_j(\zz_j)\right\|_\Omega^2 & = &  \min_{ P_j\in {\mathbb R}_{r_j}} \left\|H - \sum_{j=1}^p P_jG_j \right\|^2 \\
                                                                                                   & = & \min_{ P_j\in {\mathbb R}_{r_j}} \left\|Q_j - P_jG_j\right\|^2,
\end{eqnarray*}
where $Q_j$ and $G_j$ are as in (\ref{hpjgj1}), and $j=1,\ldots,p$. The problem $\displaystyle\min_{ P_j\in {\mathbb R}_{r_j}} \left\|Q_j - P_jG_j\right\|^2$ has been studied in  \cite{Torokhti2007,tor5277} where its solution has been obtained in form (\ref{fjhjrgj}). In Theorem  \ref{th01}, the  conditions associated with $r_j$, rank of matrix $Q_jR_{G_j}$ and the singular values of matrix $Q_jR_{G_j}$ follow from the conditions of uniqueness of the SVD (see, e.g., \cite{Torokhti2007}).

\subsubsection{Proof of Theorem \ref{th011} }\label{}

Denote $\overline{\x}=\x-\displaystyle\sum_{i=1,\atop i\neq j}^p \p_j\zz_j$. Then we have
\begin{eqnarray*}
&&\hspace{-2cm}f(P_1,...P_{j-1},\widehat{P}_j,P_{j+1},...,P_p) \\
& = & \|\x-\sum_{i=1,\atop i\neq j}^p \p_j\zz_j-\widehat{\p}_j\zz_j\|^2_{\Omega}\nonumber\\
& = &   \|\overline{\x}-\widehat{\p}_j\zz_j\|^2_{\Omega}\nonumber\\
\end{eqnarray*}
On the basis of Theorem 2 in\cite{2001309}, we obtain
\begin{eqnarray*}
\|\overline{\x}-\widehat{\p}_j\zz_j\|^2_{\Omega} & =& \text{tr}\{E_{\overline{x}\overline{x}}\}-\sum_{i=1}^{r_j}\delta_i-\text{tr}\{C_jH_j^\dagger C_j^T\}\\
                                                 & =& \text{tr}\{E_{\overline{x}\overline{x}}\}-\sum_{i=1}^{r_j}\delta_i-\sum_{i=1}^{m_j}\mu_{j,i}.
\end{eqnarray*}
Further,
\begin{eqnarray}\label{eq7}
\text{tr}\{E_{\overline{x}\overline{x}}\} & = & \text{tr}\{E_{x-w_j,x-w_j}\}\nonumber\\
                                          & = & \text{tr}\{E_{xx}\}-2\;\text{tr}\{E_{w_jx}\}+\text{tr}\{E_{w_jw_j}\}.
\end{eqnarray}
Then (\ref{errorf1p}) follows.

\subsubsection{Comparison with the Linear WSN in (\ref{f1p})}\label{comparison lin}

Recall that (\ref{errorf1p}) represents the error  associated with the second degree WSN represented by (\ref{ptsjyj}).
 For the linear WSN considered in  Section \ref{linear}, at each iteration of Algorithm 1, matrix $\widehat{P}_j$ should be replaced with matrix $\widehat{F}_j$ where $\widehat{F}_j$ is represented by the  GKLT (considered, in particular, in \cite{torbook2007}) or by its particular case given in  \cite{Brillinger2001}. We wish to compare the error in (\ref{errorf1p})   with the error associated with the linear WSN.
To this end, first, we  establish the error representation for the linear WSN.
Let us denote
$$f_L (F_1,...,F_p) = \displaystyle\left\|{\bf x}-\sum_{j=1}^p \f_j(\y_j)\right\|_\Omega^2,$$ $\f_j$ is as in (\ref{f1p}),
 $\widetilde{\w}_j=\sum_{i=1 \atop i\neq j}^p\f_i\y_i$, $\alpha_j = \text{tr}\{2\; E_{\widetilde{w}_jx} - E_{\widetilde{w}_j\widetilde{w}_j}\}$ and $\widetilde{\x}=\x- \widetilde{\w}_j$. Further, we write
$\sigma_1,...,\sigma_{r_j}$ for the first $r_j$ eigenvalues in the SVD for matrix $E_{\widetilde{x}y_j}E_{y_jy_j}^\dagger E_{y_j\widetilde{x}}$.
In the theorem that follows we consider the case when  $\widehat{F}_j$ is determined by the GKLT.
\begin{theorem}\label{th012}
 For given $F_1,\ldots,F_{j-1},$ $F_{j+1},$ $\ldots,F_p$, the error associated with the optimal matrix $\widehat{F}_j$ for the linear WSN is given by
\begin{equation}\label{errorf1p1}
f_L(F_1,...,F_{j-1},\widehat{F}_j,F_{j+1},...,F_p) =  \text{tr}\{E_{xx}\}-\sum_{i=1}^{r_j}\sigma_i-\alpha_j.
\end{equation}
\end{theorem}

 \begin{IEEEproof} On the basis of \cite{torbook2007} (pp. 313),
\begin{eqnarray*}
& & \hspace{-2cm}f_L(F_1,...,F_{j-1},\widehat{F}_j,F_{j+1},...,F_p) \\
& = & \|\x-\sum_{i=1,\atop i\neq j}^p \f_j\y_j-\widehat{\f}_j\y_j\|^2_{\Omega}\nonumber\\
& = &   \|\widetilde{\x}-\widehat{\f}_j\y_j\|^2_{\Omega}\nonumber\\
& = &   \text{tr}\{E_{\widetilde{x} \widetilde{x}}\}-\sum_{i=1}^{r_j}\sigma_i
\end{eqnarray*}
 where
 \begin{eqnarray}\label{}
\text{tr}\{E_{\widetilde{x} \widetilde{x}}\} & = & \text{tr}\{E_{x-\widetilde{w}_j,x-\widetilde{w}_j}\}\nonumber\\
                                          & = & \text{tr}\{E_{xx}\}-2\;\text{tr}\{E_{\widetilde{w}_jx}\}+\text{tr}\{E_{\widetilde{w}_j\widetilde{w}_j}\}.
\end{eqnarray}
The above implies (\ref{errorf1p1}).
 \end{IEEEproof}

The following theorem establishes a condition under which the error in (\ref{errorf1p}) associated with the second degree WSN  is less than that in  (\ref{errorf1p1}) for the linear WSN.
\begin{theorem}\label{th013}
If
\begin{eqnarray*}
\alpha_j - \beta_j < \sum_{i=1}^{r_j}(\delta_i-\sigma_i) + \sum_{i=1}^{m}\mu_{j,i}
\end{eqnarray*}
then
\begin{equation*}
f(P_1,...,P_{j-1},\widehat{P}_j,P_{j+1},...,P_p) <  f_L(F_1,...,F_{j-1},\widehat{F}_j,F_{j+1},...,F_p).
\end{equation*}

\end{theorem}

 \begin{IEEEproof}
The proof follows directly from  (\ref{errorf1p}) and (\ref{errorf1p1}).
 \end{IEEEproof}

\subsubsection{Comparison with  Linear WSNs  in \cite{4276987,4475387}}
A theoretical comparison  of the error  associated with  the second degree WSN  in (\ref{errorf1p}) and
  the errors associated with the linear WSNs studied in \cite{4276987,4475387} is  difficult because of the following. The error analysis in \cite{4276987,4475387}  is provided in terms which are quite different from those used in Theorem \ref{th011}. Second,  in \cite{4276987,4475387}, the error analysis is given under the assumption that the covariance matrices are invertible, which is not the case in Theorem \ref{th011}. At the same time, the errors associated with the methods in \cite{4276987} and \cite{4475387} might be represented in the form similar to that in (\ref{errorf1p1}) where $\sigma_i$ and $\alpha_j$ should specifically be obtained for  the methods in  \cite{4276987,4475387}. In this case, the comparison would be provided in the form similar to that in Theorem \ref{th013}.
  Other way for the numerical comparison is to use the error representation in (\ref{eq71}). The error   associated with  the linear WSN  in \cite{4276987} or \cite{4475387}, follows from  (\ref{eq71}) where $z$ should be replaced with $y$ and $P^{(q+1)}$ should be replaced with the corresponding iteration of the method \cite{4276987} or \cite{4475387}. Of course, this is a posteriori comparison, nevertheless,  it is an opportunity to obtain a numerical representation of the errors under consideration. In fact, this comparison is provided in the simulations in Section \ref{sim}.

\subsubsection{Convergence of Algorithm 1}\label{}
Convergence of the method presented in Section \ref{structure} can be shown on the basis of the results given in \cite{chen87,Zhening2015,LiLiu2012} as follows.
We call ${\bfP}=\left(P_1,...,P_p\right)\in\rt_{r_1,\ldots,r_p}$ a point in the space $\rt_{r_1,\ldots,r_p}.$ For every point  $\bfP\in\rt_{r_1,\ldots,r_p}$, define a set
$$\rt_{r_j}^{\mbox{\scriptsize\boldmath $P$}} = \left\{\left(P_1,\ldots,P_{j-1}\right)\right\} \times  \rt_{r_j} \times \left\{\left(P_{j+1},\ldots,P_{p}\right) \right\},$$
for $j=1,\ldots,p$. A coordinate-wise minimum point of the procedure represented by Algorithm 1 is denoted by
  ${\bfP}^*=\left(P_1^*,...,P_p^*\right)$ where\footnote{The RHS in (\ref{fjarg}) is a set since the solution of problem $\displaystyle\min_{P_j\in {\mathbb R}_{r_j}}\;\phi(P_1^*,...,P_{j-1}^*,P_j,P_{j+1}^*,...,P_p^*) $  is not unique.}
\begin{eqnarray}\label{fjarg}
P_j^*\in \left\{\arg\min_{P_j\in {\mathbb R}_{r_j}}\;\phi(P_1^*,...,P_{j-1}^*,P_j,P_{j+1}^*,...,P_p^*) \right\}.
\end{eqnarray}
This point is a local minimum of objective function in (\ref{p1ph1p}), $\phi({\bfP}) = \left\|H - \displaystyle\sum_{j=1}^p P_jG_j \right\|^2$.\footnote{There could be other local minimums defined differently from that in (\ref{fjarg}).}
Note that ${\bfP}^{(q+1)}$ in Algorithm 1 and $\bfP^*$ defined by (\ref{fjarg}) are, of course, different.

We also need the following auxiliary result.

\begin{lemma}\label{lem1}
The sequence $\{\bfP^{(q)} \}$ generated by Algorithm 1  is bounded.
\end{lemma}

\begin{IEEEproof}
Consider $\widetilde{\phi}(\bfP) = \widetilde{\phi}(P_1,\ldots, P_p) = \|P G - H\|$ where $G =[G_1,\ldots, G_p]$, and $G_j$, for $j=1,\ldots,p$, and $H$ are given in (\ref{xfy1p}) and (\ref{ezzg1p}). Then $\|P G - H\|\geq \|P G\| - \|H\|$, i.e.,
\begin{eqnarray}\label{pgh1}
\left\|P G\right\| - \left\|H\right\|\leq \widetilde{\phi}(P_1,\ldots, P_p).
 \end{eqnarray}
 Suppose sequence $\{\bfP^{(q)} \}$ is unbounded, i.e., $\|P^{(q)}\|\rightarrow \infty$ as $q\rightarrow \infty$. Then (\ref{pgh1}) implies
$\left\|P^{(q)} G\right\| - \left\|H\right\|\rightarrow \infty$ as $q\rightarrow \infty$. Therefore, $\widetilde{\phi}(P^{(q)}_1,\ldots, P^{(q)}_p)\rightarrow \infty$ as $q\rightarrow \infty$ and then  ${\phi}(P^{(q)}_1,\ldots, P^{(q)}_p)\rightarrow \infty$ as $q\rightarrow \infty$.  But it conflicts with Algorithm 1 where
$$
0\leq {\phi}(P^{(q)}_1,\ldots, P^{(q)}_p)  \leq {\phi}(P^{(0)}_1,\ldots, P^{(0)}_p).
$$
Thus, $\{\bfP^{(q)} \}$  is bounded.
\end{IEEEproof}

Now we are in the position to show convergence of Algorithm 1. For $\bfP^{(q)}$ defined by Algorithm 1, denote
\begin{eqnarray}\label{bfinfty}
\phi(\check{\bfP}) =\lim_{q\rightarrow \infty} \phi(\bfP^{(q)}).
\end{eqnarray}

\begin{theorem}\label{th014}
Point  $\check{\bfP} $ defined by (\ref{bfinfty}) is the coordinate-wise minimum of Algorithm 1.
\end{theorem}

\begin{IEEEproof}
For each fixed ${\bfP}=(P_1,...,P_p)$,  a so-called best response matrix to matrix $P_j$ is denoted by  $\po_j^{\mbox{\scriptsize\boldmath $P$}}$, where
$$\po_j^{\mbox{\scriptsize\boldmath $P$}}\in\left\{\arg\min_{P_j\in{\mathbb R}_{r_j}}\;\phi(P_1,...P_{j-1},P_j,P_{j+1},...,P_p)\right\}.$$
Let $\{{\bfP}^{(q)}\}$ be a sequence generated by Algorithm 1, where ${\bfP}^{(q)}=(P_1^{(q)},...,P_p^{(q)})$.
 Since each $\rt_{r_j}^{\mbox{\scriptsize\boldmath $P$}} $ is  closed  \cite[p. 304]{tu2007introduction} and sequence $\{\bfP^{(q)} \}$  is bounded, there is a subsequence $\{{\bfP}^{(q_s)}\}$ such that $(P_1^{(q_s)},...,P_p^{(q_s)})\rightarrow(P_1^{*},...,P_p^{*})={\bfP}^*$ as $s\rightarrow\infty$. Then, for any $j=1,...,p$, we have
\begin{eqnarray*}
& & \hspace{-1cm} \phi(P_1^{(q_s)},...,P_{j-1}^{(q_s)},\po_j^{{\mbox{\scriptsize\boldmath $P$}}^*},P_{j+1}^{(q_s)},...,P_p^{(q_s)})\\
& \geq & \phi(P_1^{(q_s)},...,P_{j-1}^{(q_s)},\po_j^{{{\mbox{\scriptsize\boldmath $P$}}}^{(q_s)}},P_{j+1}^{(q_s)},...,P_p^{(q_s)})\\
& \geq & \phi(P_1^{(q_s+1)},...,P_{j-1}^{(q_s+1)},P_{j}^{(q_s+1)},P_{j+1}^{(q_s+1)},...,P_p^{(q_s+1)})\\
& \geq & \phi(P_1^{(q_{s+1})},...,P_{j-1}^{(q_{s+1})},P_{j}^{(q_{s+1})},P_{j+1}^{(q_{s+1})},...,P_p^{(q_{s+1})})
\end{eqnarray*}
By continuity, when $s\rightarrow\infty$,
\begin{multline*}
\hspace{-0.2cm}\phi(P_1^{*},...,P_{j-1}^{*},\po_j^{{\mbox{\scriptsize\boldmath $P$}}^*},P_{j+1}^{*},...,P_p^{*})\geq \\ \phi(P_1^{*},...,P_{j-1}^{*},P_{j}^{*},P_{j+1}^{*},...,P_p^{*}),
\end{multline*}
which implies that above should hold as an equality, since the inequality is true by the definition of the best response matrix $\po_j^{{\mbox{\scriptsize\boldmath $P$}}^*}$. Thus, $P_j^*$ is such as in (\ref{fjarg}), i.e.  $P_j^*$ is a  solution of the problem
$$\min_{P_j\in\mathcal{R}_{r_j}}\;\phi(P_1^*,...,P_{j-1}^*,P_j,P_{j+1}^*,...,P_p^*),\;\forall j=1,...,p.$$
\end{IEEEproof}

\subsection{Justification of Algorithm 2}\label{convergence2}

For $\bfP^{(q)}$ defined by Algorithm 2, denote
\begin{eqnarray}\label{bfinfty2}
\psi({\bfP}^*) =\lim_{q\rightarrow \infty} \psi(\bfP^{(q)}).
\end{eqnarray}

\begin{theorem}\label{th015}
 Point  ${\bfP}^*=(T^*,S_1^{*},...,S_p^{*}) $ defined by (\ref{bfinfty2}) is the coordinate-wise minimum of Algorithm 2.
\end{theorem}

\begin{IEEEproof}
The proof is  similar to that for Theorem \ref{th014}. Therefore, we only provide a related sketch.
For each fixed ${\bf P}=(T,S_1,...,S_p)$, denote by $\it\Psi_j^{\bf P}$  a best response matrix to $S_j$ defined by
$$R_j^{\bf P}\in\left\{\arg\min_{S_j\in\mathbb{R}^{r_j\times n}}\;\psi(T,S_1,...S_{j-1},S_j,S_{j+1},...,S_p)\right\}.$$
By Algorithm 2, a member of sequence  $\{\bfP^{(q)} \}$ is given by ${\bf P}^{(q)}=(T^{(q)},S_1^{(q)},...,S_p^{(q)})$. Similar to Lemma \ref{lem1}, sequence $\{\bfP^{(q)} \}$ provided by Algorithm 2 is bounded. Then there is a subsequence $\{{\bf P}^{(q_v)}\}$ with ${\bf P}^{(q_v)}=(T^{(q_v)},S_1^{(q_v)},...,S_p^{(q_v)})$ such that  $(T^{(q_v)},S_1^{(q_v)},...,S_p^{(q_v)})\rightarrow(T^*,S_1^{*},...,S_p^{*})={\bf P}^*$ as $v\rightarrow\infty$. Then, for any $j=1,...,p$, we have
\begin{multline*}
\hspace{-0.25cm}\psi(T^{(q_v)},S_1^{(q_v)},...,S_{j-1}^{(q_v)},R_j^{\bf P^*},S_{j+1}^{(q_v)},...,S_p^{(q_v)}) \geq \\ \psi(T^{(q_{t+1})},S_1^{(q_{t+1})},...,S_{j-1}^{(q_{t+1})},S_{j}^{(q_{t+1})},S_{j+1}^{(q_{t+1})},...,S_p^{(q_{t+1})})
\end{multline*}
and, for $v\rightarrow\infty$,
\begin{multline*}
\hspace{-0.25cm}\psi(T^*,S_1^{*},...,S_{j-1}^{*},R_j^{\bf P^*},S_{j+1}^{*},...,S_p^{*})\geq \\ \psi(T^*,S_1^{*},...,S_{j-1}^{*},S_{j}^{*},S_{j+1}^{*},...,S_p^{*}),
\end{multline*}
which implies that above should hold as an equality. Thus, $S_j^*$ is the best response to $(S_1^*,...S_{j-1}^*,S_{j+1}^*,...,S_p^*)$, or equivalently,  $S_j^*$ is the optimal solution for the problem
$$\max_{S_j\in\mathbb{R}^{r_j\times n}}\;\psi(T^*,S_1^*,...,S_{j-1}^*,S_j,S_{j+1}^*,...,S_p^*),\;\forall j=1,...,p.$$
The proof is similar when
$$R_j^{\bf P}\in\left\{\arg\min_{T\in\mathbb{R}^{m\times r_j}}\;\psi(T,S_1,...,S_p)\right\}.$$
\end{IEEEproof}

\bibliographystyle{IEEEtran}
\bibliography{references}

\end{document}